\documentstyle[epsfig]{ar}
\leftmargini\parindent
\begin{document}
\def\openone{\leavevmode\hbox{\small1\kern-3.3pt\normalsize1}}
\def\Re{{\rm Re\:}}
\def\Im{{\rm Im\:}}
\def\Tr{{\rm Tr\:}}
\def\Str{{\rm Str\:}}
\def\diag{{\rm diag\:}}
\def\be{\begin{eqnarray}}
\def\ee{\end{eqnarray}}
\newcommand{\mat}{\left ( \begin{array}{cc}}
\newcommand{\emat}{\end{array} \right )}
\newcommand{\vect}{\left ( \begin{array}{c}}
\newcommand{\evect}{\end{array} \right )}
\title{Random Matrix Theory and Chiral Symmetry in QCD}
\markboth{J.J.M. Verbaarschot and T. Wettig}{Random Matrix Theory and
  Chiral Symmetry in QCD}

\author{J.J.M. Verbaarschot
\affiliation{Department of Physics and Astronomy, SUNY, Stony Brook,
  NY 11794-3800} 
T. Wettig
\affiliation{Department of Physics, Yale University, New Haven, CT
  06520-8120 and\\
  RIKEN BNL Research Center, Brookhaven National Laboratory, Upton, NY
  11973-5000}} 

\begin{keywords}
  effective low energy theories, finite volume partition function,
  lattice QCD
\end{keywords}

\begin{abstract}
  Random matrix theory is a powerful way to describe universal
  correlations of eigenvalues of complex systems. It also may
  serve as a schematic model for disorder in quantum systems. In this
  review, we discuss both types of applications of chiral random matrix
  theory to the QCD partition function. We show that constraints
  imposed by chiral symmetry and its spontaneous breaking determine
  the structure of low-energy effective partition functions for the
  Dirac spectrum. We thus derive exact results for the low-lying
  eigenvalues of the QCD Dirac operator.  We argue that the
  statistical properties of these eigenvalues are universal and can be
  described by a random matrix theory with the global symmetries of
  the QCD partition function.  The total number of such eigenvalues
  increases with the square root of the Euclidean four-volume. The
  spectral density for larger eigenvalues (but still well below a
  typical hadronic mass scale) also follows from the same low-energy
  effective partition function. The validity of the random matrix
  approach has been confirmed by many lattice QCD
  simulations in a wide parameter range.  Stimulated by the success of
  the chiral random matrix theory in the description of universal
  properties of the Dirac eigenvalues, the random matrix model is
  extended to nonzero temperature and chemical potential. In this way
  we obtain qualitative results for the QCD phase diagram and the
  spectrum of the QCD Dirac operator.  We discuss the nature of the
  quenched approximation and analyze quenched Dirac spectra at nonzero
  baryon density in terms of an effective partition function.
  Relations with other fields are also discussed.
\end{abstract}

\maketitle

\section{INTRODUCTION}
\label{sec1}

Well before the advent of quantum chromodynamics (QCD), the theory of
the strong force, it was realized that the essential ingredients of
the hadronic world at low energies are chiral symmetry and its
spontaneous breaking.  (see e.g.\ Refs.~\cite{chiralgel,benlee}).
Mainly through lattice QCD simulations, it has become well established
that chiral symmetry breaking by the vacuum state of QCD is a
nonperturbative phenomenon that results from the interaction of many
microscopic degrees of freedom.  We argue in this review that the
complexity of the QCD vacuum leads to a low-energy description that is
completely dictated by the global symmetries of QCD.  This
interpretation of Goldstone's theorem provides a natural duality
between a strongly interacting fundamental theory and a weakly
interacting low-energy effective theory.

\subsection{QCD and Chiral Symmetry}
\label{sec1.1}

We illustrate the concept of spontaneous symmetry breaking using the
simpler example of a classical spin system with two rotational degrees
of freedom.  The Hamiltonian of this system has a certain symmetry: It
is invariant under rotations.  In mathematical language, the symmetry
group is $G={\rm O}(3)$.  However, at low temperatures, the ground
state of the system does not exhibit this symmetry.  In a
small external magnetic field, which breaks the rotational invariance
explicitly, the spins will polarize in the direction of the magnetic
field. In the thermodynamic limit, the spins remain polarized even if
the magnetic field is switched off completely.  This phenomenon is
known as spontaneous magnetization.  The ground state is no longer
invariant under O(3) rotations but only under O(2) rotations in the
plane perpendicular to the spontaneous magnetization.  In mathematical
language, the full symmetry group $G={\rm O}(3)$ is
spontaneously broken to a smaller symmetry group $H={\rm O}(2)$. The
spontaneously broken phase is characterized by low-energy excitations
in the form of spin waves in the plane perpendicular to the
spontaneous magnetization.  This is a consequence of a general theorem
known as Goldstone's theorem \cite{Goldstone}, which tells us that
spontaneous breaking of a continuous symmetry leads to low-lying
excitations, the Goldstone modes, with a mass that vanishes in the
absence of a symmetry-breaking field.

The Goldstone modes are given by the fluctuations in the plane
perpendicular to the direction of the spontaneous magnetization. Thus,
the spin system has two Goldstone modes. In general,
spontaneous symmetry breaking in a system of spins with $n$ components
is associated with $n-1$ Goldstone modes.  This number is also equal
to the number of generators of the coset $G/H$ [the number of
generators of O($n$) is $n(n-1)/2$].

A spontaneously broken symmetry is characterized by an order
parameter, which in this case is the spontaneous magnetization.  At
nonzero temperature, the alignment of the spins is counteracted by
their thermal motion, and above a critical temperature (the Curie
temperature) the spontaneous magnetization vanishes.

Let us now consider the hadronic world and interpret the particle
spectrum in terms of the concepts discussed above.  We will look for
the simplest theory consistent with the following two empirical facts:
$(a)$ there are three bosonic particles, the pions, that are much
lighter than all other hadrons and $(b)$ the proton and the neutron
have almost the same mass.  Fact $(a)$ implies that there are three
Goldstone bosons associated with a spontaneously broken symmetry.
Assume that the underlying theory is based on an $n$-component field
without preferred directions, i.e.\ the theory is invariant under
O($n$) transformations of the fields.  Spontaneous symmetry breaking
means that the ground state has a preferred direction, leaving $n-1$
directions for the Goldstone modes.  Since there are three pions, this
suggests that $n=4$ and that the ground state of the theory should be
symmetric under O(3).  The O(3) symmetry is the familiar isospin
symmetry, which results in the near equality of the masses of the
pions and of the mass of the proton and the neutron, fact $(b)$. As
was first conjectured by Gell-Mann and L\'evy \cite{chiralgel}, the
theory of the strong force is based on an O(4) invariance
spontaneously broken to O(3) with nucleons transforming according to
SU(2). This is the familiar linear $\sigma$-model.  Since the pions
are not completely massless, a small O(4) symmetry-breaking mass term
should also be present in the theory.

The $\sigma$-model is a phenomenological model for the interactions of
pions and nucleons.  It has been very successful in explaining many
previously known empirical relations.  Our aim, however, is to
understand chiral symmetry in terms of QCD, the fundamental theory of
the strong interactions.  Quantum chromodynamics is a gauge theory of quarks
that come in six different flavors and three different colors. The
gauge field interaction is according to the non-Abelian SU(3) color
group.  Two of the six quarks are nearly massless, and at low energies
QCD is well approximated by a theory with only the two lightest
quarks. They are mixed by the SU(2) isospin symmetry group.  This
symmetry is exact for degenerate quark masses.  For massless quarks,
there is an additional symmetry.  The helicity of a particle is a good
quantum number, and the right-handed and left-handed quarks can be
rotated independently.  The isospin or vector symmetry rotates both
chiralities in the same way whereas the axial SU(2) group rotates them
in the opposite direction.  This explains that the chiral symmetry group in
the massless case is $G=\rm SU(2)\times SU(2)$, which is isomorphic to
O(4).  The mass term breaks this symmetry explicitly and thus plays
the role of the magnetic field in the spin system.  Even in the
massless case, however, the vacuum state of QCD is characterized by a
nonzero expectation value of the chiral condensate, which, like the
mass term, mixes right-handed and left-handed quarks.  Thus, in the
vacuum, the axial part of the symmetry group $G$ is broken
spontaneously.  The ground state is not unique, and the degenerate
states are connected by the broken group $G$.  The degeneracy can be
lifted by means of a small mass term, and in the thermodynamic limit
the system will be frozen in this direction, i.e.\ the direction of
the QCD vacuum state is determined by the mass term.  This is exactly
the same situation as in the spin system discussed above.  The
Goldstone modes analogous to spin waves are the pions, with a mass that
is well below the typical hadronic mass scale of about 1\,GeV.  As in
the spin system, we expect that the expectation value of the chiral
condensate will become zero above a critical temperature.  This
phenomenon is known as the restoration of chiral symmetry. The
chirally symmetric phase probably existed in the early universe and
may be produced in relativistic heavy-ion collisions.

Although QCD is the consistent theory of the strong interactions, many
questions remain unanswered. For example, why have its constituents
never been observed in nature? This phenomenon is known as confinement
and means that all physical states are color singlets.  QCD is best
understood at high energies, where, because of asymptotic freedom,
quarks and gluons become weakly interacting and perturbative
calculations are possible.  At low energies, on the other hand, it is
necessary to rely on nonperturbative approaches. One approach is to study
the QCD partition function by means of Monte Carlo simulations of a
discretized version of the QCD action. This approach has been very
fruitful, and a great deal of our understanding of low-energy QCD is
based on such calculations (for a recent review, see Ref.~\cite{detar}).
The drawback is that large-scale simulations do not necessarily
provide a simple picture of the relevant degrees of freedom.
Therefore, it is often advantageous to study QCD by means of
effective models or theories. One example is the instanton liquid
vacuum \cite{instantonl}, which is a model for an ensemble of relevant
gauge field configurations.  Effective low-energy theories are a
second example. We already mentioned the $\sigma$-model for pions and
nucleons. 
However, because of the spontaneous breaking of chiral symmetry and
the existence of a mass gap in QCD, one can do better. Based on chiral
symmetry one can formulate an exact low-energy theory for the
Goldstone modes. This nonlinear $\sigma$-model is the basis for a
systematic low-energy expansion \cite{Weinberg,GaL} that is
discussed in detail in Secs.~\ref{sec1.3} and \ref{sec:effective}.

The QCD partition function can be written as a Euclidean path integral
that can be expressed as the expectation value of the fermion
determinant,
\begin{equation}
  \label{eq1}
  Z = \Bigl\langle \prod_f \det ({\cal D}+m_f) \Bigr\rangle\:.
\end{equation}
Here the average is over all gauge fields weighted by the Euclidean
Yang-Mills action, the product is over quark flavors of mass $m_f$,
and ${\cal D}$ is the Dirac operator, which we introduce in great
detail in Sec.~\ref{sec:QCD}.  The fermion determinant can be
expressed as a product over the eigenvalues $i\lambda_n$ of the Dirac
operator.  Therefore, we may also interpret the average as an average
over the eigenvalues with probability distribution determined by the
gauge field dynamics.  We have argued that spontaneous breaking of
chiral symmetry means that a small quark mass leads to a macroscopic
realignment of the QCD vacuum.  It is clear from the QCD partition
function that this is only possible if there is an accumulation of
Dirac eigenvalues near zero. Otherwise, a small mass term would be
completely dominated by the much larger eigenvalues in the factors
of $(i\lambda_n+m_f)$.  For a free Dirac operator in a
four-dimensional box, the eigenvalue density is proportional to
$\lambda^3$ and thus vanishes near zero.  Therefore, the small
eigenvalues must be due to interactions mediated by the gauge
fields. 

There are two possibilities. First, the eigenvalues may
originate from the exactly zero Dirac eigenvalues in the field of an
instanton.  At a nonzero density of the liquid of instantons and
anti-instantons, the zero eigenvalues are distributed over a band
because of interactions that lift the degeneracy of the eigenvalues
\cite{shurmit}.  The eigenvalue repulsion results in a nonzero
eigenvalue density near zero.  Second, the eigenvalues may originate
from the bulk of the Dirac spectrum. As is the case for any
interaction in quantum mechanics, the interactions mediated by the
gauge fields lead to a repulsion of the eigenvalues. For increasing
interaction strength, it is then advantageous for the eigenvalues to
move to a region with a low eigenvalue density. Both mechanisms of
spontaneous chiral symmetry breaking rely on the repulsion of
eigenvalues in interacting systems.  This topic has been investigated
in detail by means of random matrix theory, discussed in
the next subsection.

\begin{figure}
  \begin{center}
    \epsfig{figure=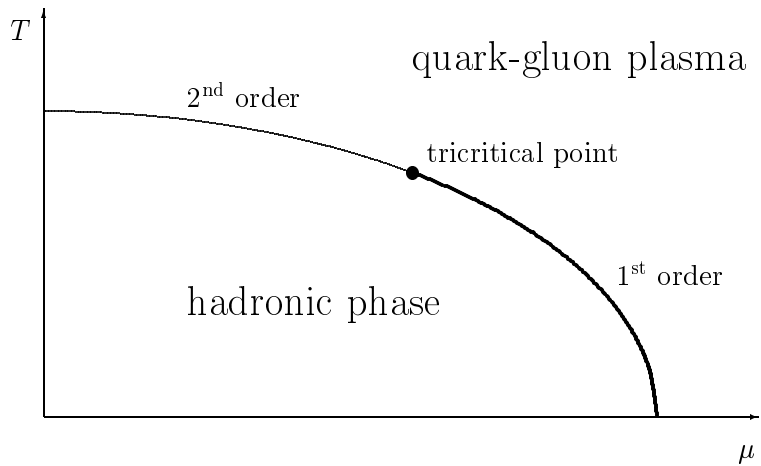,width=100mm}
    \caption{A minimal phase diagram for QCD with two massless flavors
      at nonzero temperature and density.}
    \label{fig1}
  \end{center}
\end{figure}

As mentioned above, we expect chiral symmetry to be
restored at high temperatures \cite{smilref}.  This is confirmed by
lattice QCD simulations, which show a critical temperature of about
150\,MeV \cite{karsch}.  However, there is another direction in which
the QCD partition function can be explored, namely at nonzero baryon
number density. In this domain, rigorous results are available only at
infinite baryon number density \cite{son}. This raises the question to
what extent these results can be extrapolated to physically
interesting densities.  Because of the complex phase of the fermion
determinant, Monte Carlo simulations are not possible in this case.
Therefore, it is not surprising that many of the recent developments in
QCD at nonzero density are based on the analysis of effective models
\cite{krishna,edward-thomas}.  The picture that has emerged for two
massless flavors is that a first-order chiral phase transition occurs
at zero temperature along the chemical potential
axis. Renormalization-group arguments and lattice QCD simulations
indicate that for two 
massless flavors, a second-order phase transition occurs at zero
chemical potential along the temperature axis. The expectation is that
a line of first-order transitions and a line of second-order
transitions will extend in the $\mu T$ plane and will join at the
tricritical point, as indicated in Fig.~\ref{fig1}. A
color-superconducting phase is conjectured to exist at higher
densities \cite{krishna,edward-thomas,son} but the discussion of this
phase is beyond the scope of this review.  One of the questions we
address is the robustness of this picture based on the
dynamics of the eigenvalues of the QCD Dirac operator.

\subsection{Random Matrix Theory}
\label{sec1.2}

Random matrix theory (RMT) first appeared
in the mathematical literature in 1928 \cite{Wish28} and was first
applied to physics in the context of nuclear resonances by Wigner
almost 50 years ago \cite{Wign51}.  At that time, theoretical
approaches such as the shell model had been very successful in
describing the low-lying excitations of complex nuclei.  However,
highly excited resonances, which can be observed experimentally by
neutron scattering, could not be described by the microscopic theory.
The problem is generic: For any complex
quantum system containing many degrees of freedom with complicated
dynamics, it is very hard, if not impossible, to obtain exact results
for the energy levels far above the ground state of the system.

Having acknowledged that the highly excited states cannot be predicted
individually, one can ask whether the experimental data have some
generic statistical features that can be described theoretically.
This is where RMT comes in.  Every quantum system is
described by a Hamilton operator that can be expressed in matrix
form.  For a complex quantum system such as a large nucleus, this
matrix is very complicated, and we may not even know its details.  In
this case, one approach is to assume that all interactions
that are consistent with the symmetries of the system are equally
likely.  This means replacing the elements of the Hamilton matrix by
random numbers that are uncorrelated and distributed according to the
same probability distribution.  To obtain definite results,
observables such as the level density must then be averaged over the
random matrix elements.  This defines a statistical theory of energy
levels, which is mathematically known as RMT.  An
important point is that the random matrix must have the same
symmetries as the original Hamilton matrix. A collection of early
papers on RMT can be found in the book by Porter \cite{Porter}, and
the standard reference on RMT is the book by Mehta \cite{Mehta}.

Can such an enormous
simplification of the real problem describe
empirical data? A similar puzzle occurred in the theory of critical
phenomena, in which the critical behavior does not depend on the detailed
dynamics of the theory. The reason for this simplification is the
appearance of a length scale, the correlation length, that diverges
at the critical point.  Because of the corresponding separation of
scales, it is possible to integrate out the short-wavelength
fluctuations and renormalize the theory to a fixed-point theory that
does not depend on the details of the initial theory.  What is the
separation of scales that takes place in quantum spectra?  The two
basic scales are the average level spacing and the scale of the
variation of the average level spacing. The equivalent of the
correlation length is the inverse average level spacing, which diverges
in the thermodynamic limit or the semiclassical limit. We thus expect
that spectral properties on the scale of the average level spacing do
not depend on the details of the underlying dynamics. They are
universal.

Universal properties can be studied in the simplest model of a given
universality class and, in this way, exact analytical results for the
correlation functions can be obtained.  In the case of spectral
correlations the simplest models of the universality classes are the
Gaussian RMTs, and in the past three decades many
exact results have been derived for these models.  Unfortunately, in
most cases it is not possible to prove whether a given theory belongs
to one of the RMT universality classes.  Therefore,
random matrix predictions are usually verified by comparisons with
empirical data.

Since Wigner's original proposal, universal quantities have been
identified and computed in a variety of fields including nuclear
physics, atomic and molecular physics, disordered mesoscopic systems,
quantum systems with classically chaotic analogs, two-dimensional
quantum gravity, conformal field theory, and QCD.  A recent
comprehensive review of the applications of RMT can be found in
Ref.~\cite{Guhr98}.  RMT is now an independent subfield of
mathematical physics.  It provides a unifying description of universal
statistical features of many different quantum systems and is
applicable whenever a system is sufficiently complex.

Let us raise an interesting point here.
The eigenvalues of the Hamilton matrix for a given quantum
system are the quantities of primary interest.  Instead of going
through RMT to compute the eigenvalues, an 
alternative --- and much simpler --- approach might be to postulate
random eigenvalues.  It turns out that this does not describe
empirical data, at least not if the random eigenvalues are
uncorrelated.  Therefore, in addition, one would have to postulate how
the eigenvalues are correlated, and it is not at all clear how to do
this.  In RMT, on the other hand, one starts with
uncorrelated random matrix elements.  At the end of the calculation,
one finds that the resulting eigenvalues are strongly correlated in
precisely the right way to describe the data.

In most applications, RMT is used to describe the statistics of energy
levels, i.e.\ of the eigenvalues of the Hamilton operator.  For the
Euclidean QCD partition function, it is more natural to construct a
random matrix model for the Dirac operator.  As shown in the
previous subsection, the spectrum of this operator is intimately
related to the phenomenon of chiral symmetry breaking.  This
establishes the connection between RMT and chiral symmetry announced
in the title of this review.  We use random matrix methods to
study the Dirac spectrum and its implications for chiral symmetry
breaking.

It is important to note that RMT
can only provide an exact description of universal
quantities. It cannot be used for the calculation of nonuniversal
observables such as the average level spacing. Such behavior is
well known in statistical mechanics where, for example, the Ising
model describes the critical exponents of the liquid-gas phase
transition but does not give the critical temperature.  In the case
of the distribution of the eigenvalues of a system, the global
spectral density is not described by RMT.  For
example, for the Gaussian random matrix ensembles the average spectral
density is a semicircle, whereas in real systems it typically increases
strongly with excitation energy.  In contrast, universal quantities do
not depend on the details of the dynamics of the system.  In RMT, they
do not depend on the distribution of the random matrix elements. 
It is crucial to
distinguish the universal quantities from the model-dependent ones.
In the first five sections of this review, we mainly concentrate
on the universal, model-independent features that yield exact
quantitative results.  However, it is sometimes useful to construct
random matrix models to obtain a qualitative description of the
physics in a disordered system.  As we show in Sec.~\ref{models},
schematic random matrix models for QCD at nonzero temperature and
density can yield important qualitative insights into problems that
are otherwise difficult to tackle.

\subsection{Effective Low-Energy Theories and Chiral Random Matrix Theory}
\label{sec1.3}

Let us now make the connection between QCD, effective low-energy
theories for QCD, and chiral RMT.  We concentrate here on the big
picture, postponing a more detailed discussion to
Secs.~\ref{sec:chRMT} and \ref{sec:effective}.

At low energies, quarks and gluons are confined in hadrons, i.e.\ the
particles we observe in nature are composite objects.  Instead of
attempting to describe the results of low-energy scattering
experiments in terms of quarks and gluons, it is often simpler to use
an effective theory whose elementary degrees of freedom are the
lightest particles of the theory. As mentioned earlier, in QCD the
low-lying degrees of freedom are the pions resulting from the
spontaneous breaking of chiral symmetry.  Therefore, an essential
ingredient in the construction of the effective low-energy theory is
the requirement that it correctly incorporates the chiral symmetries
of the original theory.  Since the up- and down-quark are not
completely massless in nature, the pions are not massless but have a
small mass of about 140\,MeV.  They are much lighter than
the lightest non-Goldstone particles, such as the $\rho$-meson or the
nucleons, which have a mass of about $\Lambda\sim$ 1\,GeV. This means
that at sufficiently low energies, the QCD partition function is well
approximated by the partition function of an effective low-energy
theory involving only pions.

We now consider QCD in a finite Euclidean volume $V_4=L^4$.  The
partition function is then dominated by the pions if
\begin{equation}
  \frac{1}{\Lambda}\ll L\:.
\end{equation}
This statement follows from Eq.~(\ref{trace}) below by comparing the
contribution of the pion, $\exp(-m_\pi L)$, to that of a heavier
particle, $\exp(-\Lambda L)$.  The interactions of the pions are
described by an effective chiral Lagrangian that will be given in
Sec.~\ref{sec:effective}. The fields in this Lagrangian can be
separated into zero-momentum modes (constant fields) and
nonzero-momentum modes. It was realized by Gasser and Leutwyler
\cite{GL} that there exists a kinematic regime where the fluctuations
of the zero-momentum modes dominate the fluctuations of the
nonzero-momentum modes.  This regime is given by the condition
\begin{equation}
  \label{Lmpi}
  L\ll\frac{1}{m_\pi}\:,
\end{equation}
where $m_\pi$ is the pion mass.  Intuitively, this means that the
wavelength of the pion is much larger than the linear extent of the
box.  Thus, the pion field does not vary appreciably over the size of
the box, so the derivative terms are small.  
(These statements are quantified in Sec.~\ref{sec:effective}.) We need
consider only the zero-momentum modes in the regime of
Eq.~(\ref{Lmpi}).  For constant fields, the spacetime integral in the
action can be replaced by the four-volume, and, effectively, we only
have to deal with a much simpler zero-dimensional theory.
However, the global symmetries remain important.

We are now ready to make the connection to chiral RMT.  A random
matrix, as introduced in the previous subsection, contains
independently distributed random variables.  Each element of the
matrix has the same average size.  On the other hand, the matrix
elements of the Dirac operator contain specific correlations due to
the gauge fields and the kinetic terms, which are not included in
the RMT.  Only the global symmetries of QCD are included
in the RMT. Since the
random matrix elements do not have any spacetime dependence, we
expect that in the domain of Eq.~(\ref{Lmpi}), which is dominated by constant
fields, the RMT reproduces the mass dependence of the
finite-volume partition function.

To summarize, the three ingredients in the construction of the
finite-volume effective theory from QCD in the regime of
Eq.~(\ref{Lmpi}) are 
the following: $(a)$ the global symmetries of QCD, $(b)$ the spontaneous
breaking of chiral symmetry, and $(c)$ the fact that the partition
function is dominated by the constant Goldstone fields.  There is an
almost one-to-one correspondence to the properties of the chiral
random matrix model \cite{Shur93}: $(a)$ the Dirac matrix is replaced by
a random matrix with the same global symmetries, $(b)$ chiral symmetry
is broken spontaneously in the limit of infinitely large random
matrices, and $(c)$ the random matrix elements
do not have any spacetime dependence.

\section{QCD PARTITION FUNCTION AND DIRAC SPECTRUM}
\label{sec:QCD}

After the introductory remarks on our basic philosophy in
Sec.~\ref{sec1}, we now present a detailed discussion of the
properties of the QCD partition function and the Dirac operator.

\subsection{Basic Definitions} 

The QCD partition function is defined as
\begin{equation}
  \label{trace}
  Z^{\rm QCD}=\Tr e^{-\beta H}\:,
\end{equation}
where $\beta$ is the inverse temperature and $H$ is the QCD
Hamiltonian in a box of volume $V_3=L^3$.  It can be rewritten as a
(suitably regularized) functional integral in Euclidean space,
\begin{equation}
  \label{ZQCD}
  Z^{\rm QCD}=\int DA_\mu\prod_{f=1}^{N_f}\det({\cal D}+m_f)\:
  e^{-S_{\rm YM}}\:,
\end{equation}
where $N_f$ is the number of quark flavors and $S_{\rm YM}$ is the
Euclidean Yang-Mills action.  The Euclidean four-volume is
$V_4=V_3 \beta$.  The $A_\mu$ are non-Abelian gauge fields, which can
be represented as
\begin{equation}
  A_\mu=A_\mu^a\frac{T_a}{2}\:,
\end{equation}
where the $T_a$ are the generators of the gauge group SU($N_c$).  The
number of colors is denoted by $N_c$.  The QCD Dirac operator is given
by
\begin{equation}
  \label{Dirac}
  {\cal D}=\gamma_\mu(\partial_\mu+iA_\mu)\:.
\end{equation}
This operator is anti-Hermitian, ${\cal D}^\dagger=-{\cal D}$.  The
$\gamma_\mu$ are Euclidean gamma matrices with
$\{\gamma_\mu,\gamma_\nu\} = 2\delta_{\mu\nu}$. We use the
chiral representation in which $\gamma_5\equiv\gamma_1\gamma_2\gamma_3
\gamma_4 =\diag(1,1,-1,-1)$.

\subsection{Global Symmetries}
The structure of the QCD Lagrangian is to a large extent determined by
symmetries and renormalizability.  As noted above, it is 
important to analyze these symmetries to construct the correct
effective low-energy theory and the correct random matrix model.  We
now discuss three important global symmetries of the partition
function and the Dirac operator.

\subsubsection{CHIRAL SYMMETRY AND TOPOLOGY}
\label{topology}

The Dirac operator satisfies 
\begin{equation}
  \label{anticomm}
  \{\gamma_5,{\cal D}\}=0\:.
\end{equation}
This relation is a compact expression of chiral symmetry, i.e.\ of the
fact that right-handed and left-handed quarks can be rotated
independently.  One can write down an eigenvalue equation for ${\cal
  D}$,
\begin{equation}
  \label{eveq}
  {\cal D}\psi_n=i\lambda_n\psi_n\:,
\end{equation}
where the eigenvalues and eigenfunctions depend on the gauge field in
Eq.~(\ref{Dirac}).  Using Eq.~(\ref{anticomm}) one can show that the
nonzero eigenvalues of ${\cal D}$ occur in pairs $\pm i\lambda_n$ with
eigenfunctions $\psi_n$ and $\gamma_5\psi_n$.  There can also be
eigenvalues equal to zero, $\lambda_n=0$.  The corresponding
eigenfunctions can be arranged to be simultaneous eigenfunctions of
$\gamma_5$ with eigenvalue $\pm 1$, i.e.\ these states have definite
chirality.  Denoting the number of zero eigenvalues with positive and
negative chirality by $N_+$ and $N_-$, respectively, the Atiyah-Singer
index theorem states that $\nu\equiv N_+-N_-$ is a topological
invariant that does not change under continuous changes of the gauge
field.  However, the individual numbers $N_+$ and $N_-$ are not
protected by topology, i.e.\ very small deformations of the gauge
field will lift accidental zero modes.  Thus, unless we impose very
special constraints on the gauge fields, we generically have either
$N_+=0$ or $N_-=0$.

In a chiral basis with $\gamma_5\psi^{R/L}=\pm\psi^{R/L}$, one can use
Eq.~(\ref{anticomm}) to show that $\langle\bar\psi_m^R|{\cal
  D}|\psi_n^L\rangle=0 =\langle\bar\psi_m^L|{\cal D}|\psi_n^R\rangle$
for all $m$ and $n$, where $\bar\psi=\psi^\dagger\gamma_0$.  {}From
this property and the fact that ${\cal D}$ is anti-Hermitian, it
follows that the Dirac operator has the matrix structure
\begin{equation}
  \label{block}
  {\cal D}=\left(\matrix{0&iW\cr iW^\dagger&0}\right)\:.
\end{equation}
This off-diagonal block structure is characteristic for systems with
chiral symmetry.  If there are $n+\nu$ right-handed and $n$
left-handed modes, the matrix $W$ has dimension $(n+\nu)\times n$, and
the matrix ${\cal D}$ in Eq.~(\ref{block}) has $|\nu|$ eigenvalues
equal to zero.

The QCD partition function can be decomposed into sectors of definite
topological charge $\nu$,
\begin{equation}
  \label{Ztheta}
  Z^{\rm QCD}(\theta)=
  \sum_{\nu=-\infty}^\infty e^{i\nu\theta}Z_\nu^{\rm QCD}\:.
\end{equation}
Here, the $\theta$-angle is the coefficient of the topological 
$F\tilde F$ term (which violates $P$ and $CP$ conservation) in the
Lagrangian.  We now show that the $\theta$-dependence of the partition
function can be absorbed into the phase of the quark masses.  To this
end, we introduce complex masses $m_f$ for the right-handed quarks and
the complex conjugate masses $m_f^*$ for the left-handed quarks.  The
partition function then reads
\begin{equation}
  \label{Znu}
  Z_\nu^{\rm QCD}=\Bigl\langle\prod_f\hat m_f^{|\nu|}\prod_{\lambda_n>0}
  (\lambda_n^2+m_fm_f^*)\Bigr\rangle_\nu\:,
\end{equation}
where $\hat m_f=m_f$ ($m_f^*$) for $\nu\ge0$ ($\nu<0$).  In
Eq.~(\ref{Znu}), the average is only over the gauge field
configurations with topological charge $\nu$, weighted by the
Yang-Mills action.  Since
\begin{equation}
   e^{i\nu\theta}\prod_f\hat m_f^{|\nu|}=\left\{
     \begin{array}{ll}
       \left(e^{i\theta}\prod_f m_f\right)^\nu & {\rm for}\;\nu\ge0\\
       \left[\left(e^{i\theta}\prod_f m_f\right)^*\right]^{-\nu} & 
       {\rm for}\;\nu<0\:,
     \end{array}
   \right.
\end{equation}
it follows that the $\theta$-dependence of $Z^{\rm QCD}$ is entirely
determined by the combination $e^{i\theta}\prod_f m_f$.

\subsubsection{FLAVOR SYMMETRIES}
\label{FlavorSymmetries}

The second global symmetry is flavor symmetry, whose spontaneous
breaking has profound implications for the hadron spectrum.  To make
this symmetry explicit, we rewrite the fermion determinant in $Z^{\rm
  QCD}$ as a Grassmann integral,
\begin{equation}
  \label{grassmann}
  \prod_{f=1}^{N_f}\det({\cal D}+m_f)=\int d\psi d\bar\psi\exp\Biggl[
  \int d^4x\sum_{f=1}^{N_f}\bar\psi_f({\cal D}+m_f)\psi_f\Biggr]\:,
\end{equation}
where $\bar \psi = \psi^\dagger \gamma_0$.  Again going to a chiral
basis with $\gamma_5\psi^{R/L}=\pm\psi^{R/L}$, the exponent on the
right-hand side of Eq.~(\ref{grassmann}) can be rewritten as
\begin{equation}
  \int d^4x \sum_{f=1}^{N_f}\Bigr(\bar\psi^R_f{iW^\dagger}\psi^R_f +
  \bar\psi^L_f{iW}\psi^L_f + \bar\psi^R_fm_f\psi^L_f +
  \bar\psi^L_fm_f\psi^R_f \Bigr)\:.
\end{equation}
In the chiral limit where all $m_f=0$, the fermion determinant is
invariant under the transformations
\begin{eqnarray}
\begin{array}{l@{\hspace*{10mm}}l}
    \psi^L\to U\psi^L & \bar\psi^L\to\bar\psi^LU^{-1}\\[2mm]
    \bar\psi^R\to\bar\psi^R V^{-1} & \psi^R\to V\psi^R\:.
  \end{array}  
\end{eqnarray}
The only condition on $U$ and $V$ is that their inverses must exist.
If the number of right-handed states $N_R$ is equal to the number of
left-handed states $N_L$, the symmetry is thus\footnote{Actually, the
  symmetry group is ${\rm Gl}(N_f) \times {\rm Gl}(N_f)$. The
  restriction to ${\rm U}(N_f)$ is a consequence of the Riemannian
  nature of the integration manifold (see Sec.~\ref{SpectralMass}).}
${\rm U}(N_f)\times{\rm U}(N_f)$.  However, if $N_R\ne N_L$, i.e.\ if
$\nu\ne0$, the axial symmetry group $\rm U_A(1)$ is broken explicitly
by instantons or the anomaly.  A second U(1) group, $\rm U_V(1)$,
corresponds to the conservation of baryon number.  The full flavor
symmetry group in the chiral limit is thus given by $G = {\rm
  SU}(N_f)\times {\rm SU}(N_f)$.  

If the quark masses are nonzero, the axial SU($N_f$) subgroup with
$U=V^{-1}$ is broken explicitly by the mass term.  The SU($N_f$)
vector symmetry (with $U=V$) is good for degenerate quark masses
($m_f=m$ for all $f$) but is broken explicitly for different quark
masses ($m_f\ne m_{f'}$ for all $f\ne f'$).

What is much more important than the explicit breaking, however, is
the spontaneous breaking of the axial-flavor symmetry.  For an
axial-flavor--symmetric ground state, the vacuum expectation value
$\langle\bar\psi\psi\rangle=
\langle\bar\psi^R\psi^L\rangle+\langle\bar\psi^L\psi^R\rangle$ would
be zero.  However, phenomenological arguments and lattice QCD
simulations indicate that $\langle\bar\psi\psi\rangle\approx
-(240\,{\rm MeV})^3$.  The spontaneous breaking of the axial symmetry
also follows from the absence of parity doublets and the presence of
Goldstone bosons, the pions.  The quantity
$\langle\bar\psi\psi\rangle$ is only invariant if $U=V$, i.e.\ the
vacuum state is symmetric under the flavor group $H=$ SU$_{\rm V}$($N_f$).
Thus, the vector symmetries are unbroken whereas the axial symmetries
are maximally broken.  The first of these statements is in agreement
with the Vafa-Witten theorem \cite{Vafa84}, which states that in
vector-like theories such as QCD, vector symmetries cannot be
spontaneously broken.  The reasons why the axial symmetries are
maximally broken \cite{peskin} are less well understood.  The
Goldstone manifold is $G/H={\rm SU}(N_f)$, and so there are $N_f^2-1$
Goldstone bosons.

\subsubsection{ANTI-UNITARY SYMMETRIES}
\label{anti-unitary} 

Third, we consider the anti-unitary symmetries of the Dirac operator.
According to a fundamental theorem by Wigner, a symmetry in quantum
mechanics is either unitary or anti-unitary.  An anti-unitary symmetry
operator $A$ can always be written as $A=UK$, where $U$ is a unitary
operator and $K$ is the complex conjugation operator. Below we always
consider spectra of an irreducible subspace of the unitary symmetries.
If $A=UK$ is an anti-unitary symmetry of the Dirac operator, then the
symmetry operator $A^2=(UK)^2=UU^*$ is unitary, and in an irreducible
subspace it is necessarily a multiple of the identity, $UU^* = \lambda
{\bf 1}$.  Because of this relation, $U$ and $U^*$ commute so that
$\lambda$ is real. By unitarity we have $|\lambda| = 1$, which yields
$\lambda=\pm1$.  Therefore, the anti-unitary symmetries can be
classified according to the sign of $A^2$. There are three possibilities
for the classification of Hermitian (or
anti-Hermitian) operators: $(a)$ there are no anti-unitary symmetries
(denoted by the Dyson index $\beta =2$); $(b)$ if $A^2 = 1$, it is
possible to construct a basis in which the operator is real (denoted
by the Dyson index $\beta =1$); $(c)$ if $A^2 = -1$, it is possible
to construct a basis in which the matrix elements of the operator can
be organized into real (or self-dual) quaternions (denoted by the
Dyson index $\beta=4$).

For QCD with $N_c\ge3$ and fermions in the fundamental representation
of the gauge group, there are no anti-unitary operators that commute
with the Dirac operator.  This means that the matrix $W$ in
Eq.~(\ref{block}) is a general complex matrix with no further
symmetries. There are two cases with nontrivial anti-unitary
symmetries: QCD with two colors and fermions in the fundamental
representation, and QCD with any number of colors and adjoint
fermions. We will discuss them next.

For QCD with two colors and fermions in the fundamental representation,
the Dirac operator is given by Eq.~(\ref{Dirac}) with
$A_\mu=A_\mu^a\tau_a/2$, where the $\tau_a$ are the SU(2) Pauli
matrices in color space.  The anti-unitary symmetry in this case is
\cite{Leut92,Verb94a}
\begin{equation}
  [C\tau_2K,i{\cal D}]=0 \:,
\end{equation} 
where $C=\gamma_2\gamma_4$ is the charge conjugation matrix.  The
square of the anti-unitary operator is $(C\tau_2K)^2=1$.  In this
case, it is possible to find a basis in
which the matrix $W$ in Eq.~(\ref{block}) is real for all gauge
field configurations \cite{Porter}.

As a consequence of the pseudoreality of SU(2),
the symmetry of the QCD partition function in
the chiral limit is enlarged to SU(2$N_f$)  \cite{peskin,SmV,Toub99}.
An axial U(1) is 
broken explicitly by instantons or the axial anomaly, as for $N_c\ge3$.
The chiral condensate is only invariant under an ${\rm Sp}(2N_f)$
subgroup of ${\rm SU}(2N_f)$.  The Vafa-Witten theorem, prohibiting
the spontaneous breaking of global vector symmetries and assuming
maximum breaking of the axial symmetries, thus predicts a pattern of
spontaneous chiral symmetry breaking given by SU($2N_f$) $\to$
Sp($2N_f$).  The Goldstone manifold is the coset SU($2N_f$)/Sp($2N_f$),
which is the set of antisymmetric unitary matrices. We thus have
$2N_f^2-N_f-1$ Goldstone bosons \cite{SmV,Toub99}.

For fermions in the adjoint representation of the gauge group and any
number of colors, the Dirac operator is ${\cal
  D}_{ab}=\gamma_\mu(\partial_\mu\delta_{ab}+ f_{abc}A_\mu^c)$, where
$a$ and $b$ are color indices and the $f_{abc}$ are the structure
constants of SU($N_c$).  In this case, the anti-unitary symmetry is
\cite{Leut92,Verb94a}
\begin{equation}
  [CK,i{\cal D}]=0 \:.  
\end{equation} 
Because of $(CK)^2=-1$, it follows that the eigenvalues of the Dirac
operator are twofold degenerate with linearly independent
eigenfunctions. In this case, it is possible to choose a basis in which
the matrix elements of $W$ are real (or self-dual) quaternions for
all gauge fields. The eigenvalues of such a matrix are unit
quaternions and therefore doubly degenerate \cite{Dyso62} in the
representation of the Dirac matrix as complex numbers.

Restricting ourselves to the case of even $N_f$, we can show that for
$N_f$ Majorana fermions, the
flavor symmetry group is now SU($N_f$). In this case, the chiral
condensate is only invariant under an O($N_f$) subgroup of SU($N_f$).
Applying again the Vafa-Witten theorem with maximum breaking of axial
symmetry, we expect the pattern of spontaneous chiral symmetry
breaking according to SU($N_f$) $\to$ O($N_f$), with the Goldstone
manifold given by the coset SU($N_f$)/O($N_f$). This is the set of
symmetric unitary matrices.  We thus have $(N_f+2)(N_f-1)/2$ Goldstone
bosons \cite{SmV,Toub99}.

On the lattice, the symmetries of the Dirac operator may be different
from the continuum symmetries \cite{Teper}.  We return to this
point in Sec.~\ref{LatticeTests}.

\subsection{Dirac Spectrum}
\label{diracspectrum}

Based on the eigenvalue equation (\ref{eveq}), we define the spectral
density of the Dirac operator by
\begin{equation}
  \label{rho}
  \rho(\lambda)=\Bigl\langle\sum_n\delta(\lambda-\lambda_n)
  \Bigr\rangle\:,
\end{equation}
where the average is over gauge fields weighted by the full QCD
action.  The spectral density is important because of its relation with
the order parameter for spontaneous chiral symmetry breaking, the
chiral condensate $\langle\bar\psi\psi\rangle$.  It was shown by Banks
and Casher \cite{Bank80} that
\begin{equation}
  \label{BC}
  \Sigma\equiv|\langle\bar\psi\psi\rangle|=
  \frac{\pi\rho(0)}{V}\:.
\end{equation}
To be precise, we should have written $\Sigma =
\lim_{\varepsilon\to0} \lim_{m\to0} \lim_{V\to\infty} \pi
\rho(\varepsilon)/V$, 
where it is important that the limits are taken in the order
indicated.  (In the normalization of Eq.~(\ref{rho}) the spectral
density is proportional to the volume, so the explicit factor of
$1/V$ in (\ref{BC}) is canceled to yield a finite result.)

The relation (\ref{BC}) can readily be derived.  The chiral condensate
is given by
\begin{equation}
  \label{massder}
  \langle\bar\psi\psi\rangle=-\lim_{m\to0}\lim_{V\to\infty}
  \frac{1}{VN_f}\frac{\partial}{\partial m}\log Z^{\rm QCD}(m)\:.
\end{equation}
{}From Eq.~(\ref{ZQCD}), this yields
\begin{equation}
  \label{BCsum}
  \langle\bar\psi\psi\rangle=-\lim_{m\to0}\lim_{V\to\infty}
  \left\langle\frac{1}{V}\sum_n\frac{1}{i\lambda_n+m}\right\rangle\:.
\end{equation}
Since the nonzero eigenvalues occur in pairs $\pm i\lambda_n$, their
contribution to the sum can be written as $2m/(\lambda_n^2+m^2)$ (with
$\lambda_n>0$).  For gauge fields with topological charge $\nu$, the
zero modes contribute a term $|\nu|/(mV)$.  Assuming
$\langle\nu^2\rangle\propto V$, we can drop these contributions in the
limit $V\to\infty$.  In the same limit, the sum in Eq.~(\ref{BCsum})
can be converted to an integral.  In the limit $m\to0$, we have
$2m/(\lambda^2+m^2)\to\pi\delta(\lambda)$, which yields Eq.~(\ref{BC}).
As discussed above in Sec.~\ref{sec1.1}, spontaneous chiral symmetry
breaking is encoded in an accumulation of the small Dirac eigenvalues;
for the order parameter to be nonzero, we need $\rho(0)/V>0$.

An immediate consequence of the Banks-Casher relation is that the
small eigenvalues are spaced as
\begin{equation}
  \label{spacing}
  \Delta\lambda=\frac{1}{\rho(0)}=\frac{\pi}{V\Sigma}\:,
\end{equation}
provided that $\rho(0)/V>0$.  This naturally defines a scale
\begin{equation}
  \label{scale}
  z=\lambda V\Sigma
\end{equation}
for the study of the distribution of individual eigenvalues.  For this
purpose, it is convenient to define the so-called microscopic spectral
density \cite{Shur93}
\begin{equation}
  \label{micro}
  \rho_s(z)=\lim_{V\to\infty}\frac{1}{V\Sigma}\:
  \rho\left(\frac{z}{V\Sigma}\right)\:.
\end{equation}
This function describes the extreme infrared properties of the Dirac
spectrum.  Based on the arguments in the introduction, we expect it to
be completely determined by the global symmetries of the Dirac
operator.  As we demonstrate below, $\rho_s(z)$ can be computed both from
the low-energy effective theory and from chiral RMT;
the results coincide.  Further confirmation of these ideas
will come from results of lattice QCD simulations.

At this point we would like to add some remarks on possible
ultraviolet divergences.  As an example, we consider the chiral
condensate in the $V\to\infty$ limit, but before the limit $m\to0$ is
taken.  Converting the sum in Eq.~(\ref{BCsum}) to an integral and
dropping the contribution from the zero eigenvalues, we obtain
\begin{equation}
  \label{UV}
  \langle\bar\psi\psi\rangle=-\lim_{\Lambda\to\infty}\lim_{m\to0}
  \int_0^\Lambda d\lambda\frac{2m\rho(\lambda)/V}{\lambda^2+m^2}\:,
\end{equation}
where we have introduced an ultraviolet cutoff $\Lambda$ that must be
removed 
at the end of the calculation.  Asymptotically, the spectral density
behaves as $\rho(\lambda)\sim V\lambda^3$, which means that the
integral is ultraviolet-divergent.  However, this divergence does not
contribute to $\langle\bar\psi\psi\rangle$ if the $m\to0$ limit is
taken first to yield the usual Banks-Casher relation.  Alternatively,
we can first subtract the divergent contributions to the integral
(\ref{UV}) and remove the cutoff before taking the chiral limit.

A slightly different situation arises if we consider the dependence of
the chiral condensate on a valence (or spectral) mass $m_v$ that does
not appear in the average, so that $\rho(\lambda)$ is independent of
$m_v$.  The quantity
\begin{equation}
  \label{UVSigma}
  \Sigma(m_v)=\lim_{\Lambda\to\infty}\int_0^\Lambda d\lambda
  \frac{2m_v\rho(\lambda)/V}{\lambda^2+m_v^2}
\end{equation}
is subject to ultraviolet divergences.  Using again the fact that
$\rho(\lambda)\sim V\lambda^3$ for large $\lambda$, we find that the
leading divergence is $\sim m_v\Lambda^2$.  If we consider valence
masses on the microscopic scale (\ref{scale}) we have
$m_v\sim1/(V\Sigma)$, and if the $V\to\infty$ limit is taken before
the $\Lambda\to\infty$ limit, the ultraviolet divergences are removed.

In lattice QCD simulations with lattice spacing $a$, the cutoff is
$\Lambda=1/a$. In the continuum limit, $ a\rightarrow 0$, both the
coupling constant and the $\bar \psi \psi$-operator have to be
renormalized. For a discussion of the ultraviolet divergences in a recent
lattice study of $\Sigma(m_v)$ at finite volume, finite quark mass,
and finite lattice spacing, see Ref.~\cite{Hernandezmv}.

\section{CHIRAL RANDOM MATRIX THEORY}
\label{sec:chRMT}

\subsection{Introduction of the Model}

In this section we introduce a chiral random matrix theory
(chRMT) with the global symmetries of the QCD Dirac operator.  In the
spirit of the invariant random matrix ensembles, we construct a model
with eigenfunctions distributed uniformly over the unitary unit
sphere. This is achieved by choosing Gaussian-distributed random
matrix elements. We thus arrive at the following chRMT
\cite{Shur93,Verb94a}:
\begin{equation}
  Z_{N_f,\nu}^\beta(m_1,\ldots, m_{N_f}) = 
  \int DW \prod_{f= 1}^{N_f} \det({\cal D} +m_f)
  e^{-\frac{N \beta}4 \Tr v(W^\dagger W)}\:,
\label{zrandom1}
\end{equation}
where $\beta$ is the Dyson index,
\begin{equation}
  {\cal D} = \left (\begin{array}{cc} 0 & iW\\
      iW^\dagger & 0 \end{array} \right )\:,
\end{equation}
and $W$ is an $n\times m$ matrix with $\nu = m-n$ and $N= n+m$. The
interpretation of this model is that $N$ low-lying modes interact via
a random interaction. A natural representation of this model is in the
form of gauge field configurations given by a liquid of instantons.
Then the low-lying modes are the zero modes of each instanton.  We
assume that $\nu$ does not exceed $\sqrt N$ so that, to a good
approximation, $n = N/2$ for large $N$. The parameter $N$ is
identified as the dimensionless volume of spacetime. For the
formulation of this model with explicit factors of $N/V$ included, see
Ref.~\cite{smilran}. The potential $v$ is defined by
\begin{equation}
  v(\phi) = \sum_{k\ge 1} a_k \phi^k\:.
\end{equation}
The simplest case is the Gaussian case, where $v(\phi) = \Sigma^2\phi$.
It can be shown (see Sec.~\ref{UniversalityProofs}) that the
microscopic spectral density does not depend on the higher-order terms
in this potential provided that the average spectral density near zero
remains nonzero.  The matrix elements of $W$ are either real [$\beta =
1$, chiral Gaussian Orthogonal Ensemble (chGOE)], complex [$\beta =
2$, chiral Gaussian Unitary Ensemble (chGUE)], or quaternion real
[$\beta = 4$, chiral Gaussian Symplectic Ensemble (chGSE)]. In the
latter case, the eigenvalues of ${\cal D}$ are doubly degenerate, and
the use of Majorana fermions is implemented by replacing the
determinant by its square root.  For a non-Gaussian potential
$v(\phi)$, we will omit the G in the abbreviations and use chOE, chUE,
and chSE, respectively.  Two earlier
attempts to describe QCD Dirac eigenvalues used the Wigner-Dyson
ensembles instead of the above chiral ensembles \cite{early}.

This model reproduces the following symmetries of the QCD partition
function:
\begin{itemize}
\item The $\rm U_A(1)$ symmetry. All eigenvalues of the random matrix
  Dirac operator occur in pairs $\pm i\lambda_n$ or are zero.
\item The topological structure of the QCD partition function. The
  Dirac matrix has exactly $|\nu|=|n-m|$ zero eigenvalues. This
  identifies $\nu$ as the topological sector of the model.
\item The flavor symmetry, which is the same as in QCD. For $\beta = 2$ it is
  ${\rm SU}(N_f) \times {\rm SU}(N_f)$, for $\beta = 1$ it is ${\rm
    SU}(2N_f)$, and for $\beta = 4$ it is ${\rm SU}(N_f)$ (each
  Majorana flavor counts as 1/2 Dirac flavor).
\item The chiral symmetry, which is broken spontaneously with a chiral
  condensate given by
  \begin{equation}                                                 
    \Sigma = \lim_{N\rightarrow \infty} {\pi \rho(0)}/N\:.
  \end{equation}
  ($N$ is interpreted as the dimensionless volume of spacetime.) The
  symmetry-breaking pattern is {\cite{SmV}} ${\rm SU}(N_f) \times {\rm
    SU}(N_f)\to{\rm SU}(N_f)$, ${\rm SU}(2N_f)\to{\rm Sp}(2N_f)$, and
  ${\rm SU}(N_f)\to{\rm O}(N_f)$ for $\beta = 2$, 1, and 4,
  respectively --- the same as in QCD \cite{peskin}.
\item The anti-unitary symmetries.  These are implemented by choosing
  the matrix elements of $W$ to be real, complex, or quaternion real
  for $\beta =1$, $\beta =2$, and $ \beta =4$, respectively.
\end{itemize}
Along with the invariant random matrix ensembles, the chiral
ensembles are part of a larger classification scheme that also
includes ensembles for the description of disordered superconductors
\cite{Altland}. In total, 10 different families of random matrix
ensembles have been identified. They correspond one-to-one 
with the Cartan classification of symmetric spaces \cite{class}.

The uniform distribution of the eigenfunctions over the unitary unit
sphere is expressed as the invariance
\begin{equation}
  W \rightarrow U^\dagger W V\:,
\label{inv}
\end{equation}
where the $n\times n$ matrix $U$ and the $m\times m$ matrix $V$ are
orthogonal matrices for $\beta=1$, unitary matrices for $\beta = 2$,
and symplectic matrices for $\beta = 4$. This invariance makes it
possible to express the partition function in terms of eigenvalues of
$W$ defined by
\begin{equation}
  W = U^\dagger \Lambda V\:.
\end{equation}
Here, $\Lambda$ is a diagonal matrix with real diagonal matrix
elements $\lambda_k \ge 0$.  In terms of the eigenvalues, the partition
function (\ref{zrandom1}) is given by
\begin{equation}
  Z_{N_f, \nu}^{\beta}(m_1, \ldots, m_{N_f}) = 
  \int d\lambda |\Delta(\lambda^2)|^\beta
  \prod_k \lambda_k^{\beta|\nu|+\beta-1} 
  e^{-\frac{N\beta}4 v(\lambda_k^2)}
  \prod_f m_f^{|\nu|}(\lambda_k^2 + m^2_f)\:,
  \label{zeig}
\end{equation}
where the Vandermonde determinant is defined by
\begin{equation}
  \Delta(\lambda^2) = \prod_{k<l} (\lambda_k^2-\lambda_l^2)\:.
  \label{vandermonde}
\end{equation}
In Eq.~(\ref{zeig}) and elsewhere in this review, we have omitted the
normalization constant of the partition function.  {}From the joint
eigenvalue distribution (which is the integrand of Eq.~(\ref{zeig})),
we see that a nonzero topological charge $\nu$ can be introduced by
adding $\beta|\nu|/2$ massless flavors to the theory with $\nu=0$.
Therefore, this duality between flavor and topology is a general
feature of all correlation functions.  For a discussion of this
duality in terms of finite volume partition functions, see
Refs.~\cite{cam97,wadati}. 

\subsection{Sigma Model Representation of Chiral Random Matrix Models}

The chiral random matrix partition function can be evaluated using
standard random matrix methods. For simplicity, let us consider the
case $\beta=2$.  The fermion determinant can be written as a Grassmann
integral, and averaging over the Gaussian distribution function
results in a four-fermion interaction. Because of the underlying
unitary invariance (\ref{inv}) of the chRMT, the
fermionic variables only appear in invariant combinations of the form
$\sigma^{fg} \sim \bar\psi^f_i \psi^g_i$, and the partition function
can be rewritten identically in terms of these variables as
\cite{Shur93,HVeff}
\begin{equation}
  Z_{\nu}(m) = \int D \sigma \: e^{-{n\Sigma^2} \Tr
  \sigma \sigma^\dagger}{\det}^{\nu} (\sigma+m{\bf 1} ){\det}^{n}
  [(\sigma+m{\bf 1} )(\sigma^\dagger +m{\bf 1})]\:,
  \label{z2sigma}
\end{equation}
where the quark masses have been taken real and degenerate for
simplicity.  The integration is over the real and imaginary parts of
the $N_f \times N_f$ arbitrary complex matrix $\sigma$. For $m = 0$
this integral is invariant under $\sigma \rightarrow U \sigma V^{-1}$
with $U$ and $V\in{\rm SU}(N_f)$.  Let us calculate the integrals by a
saddle-point approximation.  The saddle-point equation reads (notice
that $\nu \ll n$)
\begin{equation}
  \Sigma^2 (\sigma^\dagger+m) \sigma = 1\:.
\end{equation}
The condensate at the saddle point is given by
\begin{equation}
  |\langle\bar\psi\psi\rangle| = \frac 1{2nN_f} \partial_m \log Z = 
  \frac {\Sigma^2}{2N_f}\langle {\rm Tr}(\sigma +\sigma^\dagger)\rangle \:,
\end{equation}
where the expectation value is with respect to the partition function
(\ref{z2sigma}).  The solution of the saddle-point equation,
$\bar\sigma$, is proportional to the identity matrix.  In the chiral
limit, we then find $\bar \sigma = 1/\Sigma$, and we can identify
$\Sigma$ as the chiral condensate.

The fluctuations about the saddle point can be separated into massive
modes with a curvature of order $n$ and Goldstone modes with a
curvature of order $mn$.  In the limit $n\rightarrow \infty$, the
partition function (\ref{z2sigma}) can be simplified by keeping the
integrals over the Goldstone manifold and performing the remaining
integrals by a saddle-point integration. This results in the partition
function
\begin{equation}
  \label{Znum}
  Z_{\nu}( m) = \int_{U \in {\rm U}(N_f)} DU\: {\det}^\nu U\:
  e^{ n\Sigma  \Tr (MU + M^\dagger U^{-1})}\:,
\end{equation}
which is the familiar finite volume partition function to be discussed
in Sec.~\ref{Finite-Volume}.

\subsection{$\theta$-Dependence of the Partition Function}

The $\theta$-dependence of the partition function is obtained by
summing over all topological sectors of the partition function
according to Eq.~(\ref{Ztheta}), which applies to full QCD where the
partition function is obtained by integrating over the gauge fields.
The terms $Z_\nu$ in this equation reflect the probability of
encountering gauge fields with topological charge $\nu$.

For the effective theory, the partition function in the topological
sector $\nu$ is given in Eq.~(\ref{Znum}).  However,
this $\nu$-dependence is only due to the fermion
determinant.  In the sum over $\nu$, we should therefore add weight
factors $P(\nu)$ that take into account the distribution of
topological charge of the quenched gauge fields (i.e.\ gauge fields
generated without the fermion determinant).  This yields
\begin{equation}
  Z^{\rm eff}(\theta,m) = \sum_{\nu}P(\nu) e^{i\nu \theta} Z^{\rm
  eff}_\nu(m)\:. 
\label{ztheta}
\end{equation}
Below, we argue that the weight factors can be ignored for light
quarks but that they are necessary when one considers the quenched
theory (which corresponds to $N_f=0$ or, equivalently, to the limit of
very heavy quarks).

Our starting point is that the topological susceptibility of the
quenched gauge fields is nonvanishing, i.e.
\begin{equation}
  \frac { \langle \nu^2 \rangle_q} V \ne 0\:.
\end{equation}
Invoking the central limit theorem, we assume that $P(\nu)$ is Gaussian,
\begin{equation}
  P(\nu) = \frac 1{\sqrt{2 \pi \langle \nu^2 \rangle_q}}\:
  e^{-\frac{\nu^2}{2 \langle \nu^2 \rangle_q }}\:,
\end{equation}
which has been verified by quenched lattice QCD simulations.

We analyze the partition function (\ref{ztheta}) for
$Z_{\nu}^{\rm eff}(m)$ given by the finite-volume partition function
in Eq.~(\ref{Znum}).  Replacing the sum over $\nu$ by an integral, we
obtain after integration
\begin{equation}
  Z^{\rm eff}(\theta,m)=\int_{U\in{\rm U}(N_f)} DU
  e^{-\frac{ \langle \nu^2 \rangle_q}2
    (\theta -i\Tr\log U )^2 +  \frac12\Sigma V \Tr (MU + M^\dagger
     U^{-1})}\:. 
\label{zlargen}
\end{equation}
The exponent in the integrand corresponds to the effective static
potential of QCD with a large number of colors \cite{Witten}, and
therefore the chRMT partition function reproduces all identities that
have been derived from this effective potential \cite{Shur93,janu1}.
For example, the topological charge is screened by light quarks. For a
careful analysis of the periodicity requirements of the
$\theta$-dependence of the effective partition function
(\ref{zlargen}), see Ref.~\cite{Zhit}.

Let us illustrate the screening of topological charge for $N_f = 1$
and real $m$, for which
\begin{equation}
  Z^{\rm eff}(\theta,m) = \int d \phi \: e^{-\frac{\langle \nu^2 \rangle_q}2
    (\theta -\phi )^2 + \Sigma Vm \cos \phi}\:.
\end{equation}
The integral over $\phi$ can be performed by a saddle-point
approximation resulting in \cite{Leut92}
\begin{equation}
  \label{Zmtheta}
  Z^{\rm eff}(\theta,m) = e^{mV\Sigma \cos\theta}.
\end{equation}
The topological susceptibility is given by
\begin{equation}
  \langle \nu^2\rangle = -\frac{\partial^2}{\partial{\theta^2}}
  \log Z =  m V\Sigma \qquad ({\rm for}\;\theta=0)
  \label{susm}
\end{equation}
so that the topological charge is completely screened in the chiral
limit.  This also shows that the results are insensitive to our choice
of the distribution function $P(\nu)$.  For example, in an instanton
liquid interpretation of the chiral random matrix model, the natural
choice for $P(\nu)$ is the binomial distribution $B(N-\nu,\nu)$,
resulting in the same effective potential \cite{Shur93}.

In fact, it can be concluded from Eq.~(\ref{susm}) that for light
quarks the distribution $P(\nu)$ can be ignored altogether. Let us
again illustrate this for $N_f =1$, for which the integral in
Eq.~(\ref{Znum}) results in the finite volume partition function
\begin{equation}
  Z^{\rm eff}_\nu(m) = I_\nu(mV\Sigma)\:,
\end{equation}
where $I_\nu$ is a modified Bessel function.  Summing over $\nu$ according
to Eq.~(\ref{Ztheta}), i.e.\ without the weight factors $P(\nu)$, we
obtain \cite{Damtop}
\begin{equation}
  Z^{\rm eff}(\theta, m) =\sum_{\nu}
  e^{i\nu \theta} I_\nu(mV\Sigma) = e^{mV \Sigma \cos \theta}\:,
\end{equation}
where we have used a summation formula for modified Bessel functions.
This is exactly the same result as in Eq.~(\ref{Zmtheta}). For an
extension of these results to more than one flavor, see
Ref.~\cite{Damtop}.

\subsection{Spectral Correlation Functions}
\label{sec3.1}

In this section, we discuss the spectral correlation functions and the
microscopic spectral density corresponding to the chRMT partition
function. The global spectral density is a semicircle. It is not
universal and we do not expect to find it in realistic physical
systems.  The universal quantities are local or microscopic spectral
correlation functions.  In general, a $k$-point spectral correlation
function is defined by \cite{Dyso62}
\begin{equation}
  \label{kpoint}
  R_k(\lambda_1,\ldots,\lambda_k)=\left\langle\sum_{\scriptsize
      \matrix{j_1\cdots j_k=1\cr j_p\ne j_q}}^N
    \delta(\lambda_1-E_{j_1})\cdots\delta(\lambda_k-E_{j_k})\right\rangle\:,
\end{equation}
where the $E_{j_p}$ are eigenvalues.  The quantity $R_k$ represents
the probability density to find $k$ eigenvalues, regardless of
labeling, at $\lambda_1,\ldots,\lambda_k$.  In particular,
$R_1(\lambda)=\rho(\lambda)$.  By ``local'' we mean that the energy
differences $|\lambda_p-\lambda_q|$ are of the order of a few mean
level spacings.

Because of the $\rm U_A(1)$ symmetry of QCD, all nonzero eigenvalues
come in pairs $\pm i\lambda_n$ (see Sec.~\ref{topology}).  This implies
that the origin, $\lambda=0$, is a special point of the spectrum and
that we have to consider the $R_k$ separately in the bulk of the
spectrum and near $\lambda=0$.  The latter is called the microscopic
region or the ``hard edge'' of the spectrum.\footnote{The reason for
  this nomenclature is that in a Fermi-gas formulation of RMT, where
  the eigenvalues are interpreted as the positions of particles, this
  symmetry corresponds to choosing a potential that is infinite for
  negative values of the eigenvalues. In this picture, the soft edge
  corresponds to a potential with a finite slope.}  In addition,
there is the tail or ``soft edge'' of the spectrum, which we do
not discuss. The bulk of the
spectrum is the middle region, far from either edge.

Because the random matrix partition function is known in terms of an
integral over the eigenvalues of the Dirac operator, see
Eq.~(\ref{zeig}), it is possible to obtain the spectral density and
all spectral $k$-point correlation functions by integration over $N-k$
eigenvalues. In the present case, the integrals can be performed most
conveniently by the orthogonal polynomial method. We mention only the
most important results in the sections below.

\subsubsection{BULK CORRELATIONS}

Our conventions are such that if we have $N$ eigenvalues, the global
spectral density is normalized to $N$ so that the mean level spacing
is of order $1/N$.  Local spectral correlation functions
$R_k(\lambda_1,\ldots,\lambda_k)$ in the bulk of the spectrum are thus
characterized by $\lambda_p\sim{\cal O}(1)$ and
$|\lambda_p-\lambda_q|\sim{\cal O}(1/N)$.  Universal correlation
functions are obtained by rescaling the eigenvalues according to the
average local level spacing. This procedure, known as
unfolding, is discussed in more detail in Sec.~\ref{unfolding}.
The rescaled eigenvalues with average level spacing equal to unity are
denoted by $x_k$.

In the bulk of the spectrum, the RMT results for the
$R_k$ are identical for the chiral ensembles and the corresponding
nonchiral ensembles \cite{Fox64,nagao}.  For simplicity, let us
consider the simplest ensemble, the chUE.  On the unfolded scale, we
obtain in the limit $N\to\infty$
\begin{equation}
  \label{det}
  R_k(x_1,\ldots,x_k)=\det[K(x_p,x_q)]_{p,q=1,\ldots,k}
\end{equation}
with the sine kernel
\begin{equation}
  K(x,y)=\frac{\sin\pi(x-y)}{\pi(x-y)}\:.
\end{equation}
The fact that $K(x,y)$ depends only on $|x-y|$ reflects the
translational invariance of the spectral properties after unfolding.
The results for the chOE and the chSE are somewhat more complicated
than those for the chUE \cite{Mehta}.

Other quantities can be derived from the $R_k$.  As a measure of
short-range correlations between the eigenvalues, one considers the
nearest-neighbor spacing distribution $P(s)$, which is the probability
density to have a spacing of $s$ between adjacent levels.  In random
matrix theories, one finds that $P(s) \sim s^\beta$ for small values of
$s$ and has a Gaussian tail for large spacings.  Long-range spectral
correlations are characterized by the level-number variance
$\Sigma^2(L)$ given by
\begin{equation}
  \Sigma^2(L)=\left\langle(n(L)-L)^2\right\rangle\:,
\end{equation}
where $n(L)$ is the number of levels in an interval of length $L$, or
by the spectral rigidity $\Delta_3(L)$ \cite{deltaDM}, which is defined
as an integral transform of $\Sigma^2(L)$,
\begin{equation}
  \Delta_{3}(L)=\frac{2}{L^4}\int_0^Ldu\:(L-u)(L^2-Lu-u^2)\Sigma^2(u)\:.
\end{equation}
The advantage of using the $\Delta_3$-statistic is that its
statistical fluctuations are much smaller than those of the
$\Sigma^2$-statistic.  The quantities $\Sigma^2$ and $\Delta_3$ can be
derived from the two-point function $R_2$.  To compute $P(s)$, one
needs all $k$-point functions.  These quantities are very different
for systems with correlated eigenvalues (described by RMT) and systems
with uncorrelated eigenvalues (described by the Poisson ensemble).
For example, the large-$L$ behavior of the chUE results is given by
$\Sigma^2(L)\sim(\ln L)/\pi^2$ and $\Delta_3(L)\sim(\ln L)/2\pi^2$,
whereas for uncorrelated eigenvalues one finds $\Sigma^2(L)=L$ and
$\Delta_3(L)=L/15$.

\subsubsection{MICROSCOPIC CORRELATIONS}

In the context of QCD, the small eigenvalues are more interesting than
those in the bulk because of their relation to spontaneous chiral
symmetry breaking via the Banks-Casher relation.  The local spectral
correlation functions at the hard edge of the spectrum are
characterized by $\lambda_p\sim{\cal O}(1/N)$ and
$|\lambda_p-\lambda_q|\sim{\cal O}(1/N)$.  Unfolding in the
microscopic region is easy; the eigenvalues are simply rescaled by the
mean level spacing at $\lambda=0$, which is given by
Eq.~(\ref{spacing}).  By convention, we define the unfolded
variables by $x_p=\lambda_p V\Sigma$, i.e.\ the factor of $\pi$ is
omitted.  Note that $V\propto N$.  Thus, in the unfolded variables we
have $x_p\sim{\cal O}(1)$ and $|x_p-x_q|\sim{\cal O}(1)$.  As
advertised above, the functional form of the $R_k$ is different in
the microscopic region.  The $R_k$ are still given by the determinant
of a kernel according to Eq.~(\ref{det}), but the sine kernel is now
replaced by the Bessel kernel \cite{Verb93},
\begin{equation}
  \label{Bessel}
  K(x,y)=\sqrt{xy}\:\frac{xJ_{\alpha+1}(x)J_\alpha(y)
    -yJ_\alpha(x)J_{\alpha+1}(y)}{x^2-y^2}\:,
\end{equation}
where $J_\alpha$ denotes Bessel functions and $\alpha=N_f+|\nu|$,
with $N_f$ 
the number of massless flavors and $\nu$ the topological charge.  A
mathematical discussion of the Bessel kernel can be found in Ref.
\cite{Tracy}.
  
The RMT results for the chOE and the chSE are more complicated
\cite{Verb94,Naga95,Brez96,Altland}.  Note that, in contrast to the
situation in the bulk, the microscopic correlations depend on $N_f$
and $\nu$.

The one-point function is universal in
the microscopic region.  For the chUE, the microscopic spectral
density, defined above in Eq.~(\ref{micro}), is
related to the kernel by $\rho_s(z)=K(z,z)$ and can be expressed as
\cite{Verb93,Verb94c}
\begin{equation}
  \rho_s(z)  = \frac {z}{2} \left[J^2_{N_f+|\nu|}(z) -J_{N_f+|\nu|+1}(z)
  J_{N_f+|\nu|-1}(z)\right]\:.
  \label{micro2} 
\end{equation}
It is also the generating function for the Leutwyler-Smilga sum rules,
which are essentially inverse moments of $\rho_s$ (or of higher-order
spectral correlation functions). They are discussed below.  The
results (\ref{Bessel}) and (\ref{micro2}) were first derived using
chiral RMT.  In the meantime, Eq.~(\ref{micro2}) has been confirmed by
an explicit calculation starting from a partially quenched chiral
Lagrangian \cite{OTV,Damg99}, which is discussed in the next section.  The
microscopic spectral density and the two-point correlation function
for the case $N_f+|\nu|=0$ have also been derived by means of the
supersymmetric method of RMT \cite{Andreev}.

There is a very interesting connection between the Bessel kernel and
the effective partition function.  In addition to the partition
function $Z^{(N_f)}$ for $N_f$ massless sea quarks, one can also
compute the partition function $Z^{(N_f+2)}$ for $N_f$ massless quarks
and two additional flavors with imaginary masses.  One then finds the
relation \cite{Damg98a}
\begin{equation}
  \label{massiveK}
  K(x,y)=\frac{1}{2}(xy)^{N_f+1/2}\:\frac{Z^{(N_f+2)}(ix,iy)}{Z^{(N_f)}}\:,
\end{equation}
which can also be generalized to massive sea quarks.  Thus, to obtain the
$R_k$ for $N_f$ flavors, one needs to know the partition function
for two additional flavors with imaginary masses.  This result has a
natural interpretation in terms of the effective spectral partition
function (discussed below in Sec.~\ref{sec:effective}).  The two
additional flavors are the spectral quark and its bosonic superpartner
\cite{OTV}.  The relation (\ref{massiveK}) once again demonstrates the
equivalence of the random matrix formulation and the effective field
theory in the zero mode domain. Additional consistency conditions for
the finite volume partition function are discussed in Ref.
\cite{Akeconsist}.

\subsection{Universality Proofs}
\label{UniversalityProofs}

In most random matrix calculations, the probability distribution of
the random matrix elements is assumed to be Gaussian
to simplify the calculation.  Nonuniversal
features such as the global spectral density depend on the choice of
the probability distribution.  On the other hand, universal results of
RMT do not depend on this choice, nor on other deformations of the
random matrix model.  In addition to numerous empirical verifications
of universal behavior in RMT, analytical proofs of universality have
recently been constructed for two different types of deformations.

First, consider deformations that do not break the invariance of the
theory under unitary transformations of the random matrix.  In this
case, the Gaussian distribution $P(W)\propto\exp(-Na_1\,\Tr
WW^\dagger)$ is replaced by the more general distribution
$P(W)\propto\exp(-N\sum_{k=1}^\infty a_k\,\Tr(WW^\dagger)^k)$, i.e.\
the quadratic term in the exponent is replaced by an arbitrary
polynomial. The advantage of the unitary invariance is that the
probability distribution depends only on the eigenvalues of
$WW^\dagger$.  For the chiral ensembles, it has been shown that both
the bulk and the microscopic spectral correlations remain unchanged on
the unfolded scale \cite{Brez96,Akem97,Kanz97,Nish98,massuni,feinrec}.
The weight function of QCD also contains the fermion determinant, and
it has been shown that the universal results are unchanged if the
random Dirac matrix in the determinant is replaced by a polynomial of
that matrix \cite{Spli99}.  Universal microscopic correlations have
also been found for the so-called Ginsparg-Wilson Dirac operator
\cite{Splittorff}.  Along with work on universality for the much
more widely studied Wigner-Dyson ensembles \cite{Bowick,AJ,Brez93,Been93,%
  Hack95,Frei96,kanuni,Kanz97,ake,hi,Zinn97,Eyna97}, these findings
make it 
clear that spectral correlations on the scale of the average
level spacing are strongly universal, i.e.\ they do not change 
despite substantial variations of the average spectral density. 
There have also been several interesting results on the
validity of universality for wide correlators, i.e.\ on macroscopic
scales \cite{AJ,Brez93,Been93,akewide}.

Second, consider deformations that do break the unitary invariance
by adding an arbitrary deterministic matrix $Y$ to the
random matrix $W$ in the Dirac operator.  As we discuss in
Sec.~\ref{models}, such a model can provide a schematic description of
QCD at nonzero temperature.  In particular, the matrix $Y$ can be
chosen in such a way that $\rho(0)/V$ vanishes so that chiral symmetry
is restored.  It has been proven that the bulk correlations on the
unfolded scale are unaffected by the matrix $Y$, and that the same is
true for the microscopic spectral correlations as long as $\rho(0)/V$
remains nonzero
\cite{senert,Jack96,Guhr97,Brez97,Seif99,paul-zinnjustin}.

Most universality proofs have been performed for random matrix
ensembles with $\beta=2$.  However, it is possible to establish
relations between the $\beta=2$ kernel and the kernels of the other
two ensembles with $\beta=1$ and $\beta=4$
\cite{Verb94,brezinneuberger,Sene98,Wido98,Klein00}.  {}From these
relations 
and the universality of the $\beta=2$ results, one can then infer the
universality of the $\beta=1$ and $\beta=4$ results. Recently,
interesting connections between massive correlators and the massless
correlators for $\beta =1$ and $\beta= 4$ have been made
\cite{akemann-kanzieper}, and the transition between these
ensembles and the $\beta=2$ ensemble has been studied
\cite{forrester14}.

The microscopic spectral correlations are universally given by RMT
only if $\rho(0)/V>0$.  An interesting question 
\cite{Akem98} is what happens to the $R_k$ if there is some phase
transition so that $\rho(0)/V$ vanishes.  This transition could be
triggered by, for example, nonzero temperature or a large number of flavors.
Beyond the transition point, there will be a gap in the spectrum, and
the smallest eigenvalues beyond this gap are presumably described by
the soft-edge results of RMT.  What is more interesting, however, is
the transition point at which $\rho(0)/V$  becomes zero so that
there is no gap.  This situation has been investigated in two
different ways.  First, keeping the unitary invariance, one can
fine-tune the polynomial in the exponent of $P(W)$ to make $\rho(0)/V$
just vanish \cite{Akem98}.  Second, breaking the unitary invariance,
one can choose a ``critical'' deterministic matrix $Y$ so that
$\rho(0)/V=0$ without a gap \cite{Hikami-Brezin,Jani98a}.  In both
cases, one obtains new functional forms for the microscopic spectral
correlations and a different scaling with the volume (or matrix
dimension).  However, the results are not unique; they depend on the
way one chooses to approach the $\rho(0)/V=0$ limit.  It is,
therefore, an open question as to whether the microscopic spectral
correlations of the QCD Dirac operator at the chiral phase transition
agree with one of the results obtained in a random matrix model.

\section{EFFECTIVE THEORIES AT LOW ENERGIES}
\label{sec:effective}

\subsection{Finite-Volume Partition Function}
\label{Finite-Volume}
As we have seen before, chiral symmetry is spontaneously broken by the
QCD vacuum.  According to Goldstone's theorem, this leads to the
appearance of massless modes. The Lagrangian describing the dynamics
of these modes can be obtained solely from the symmetries of the
QCD partition function.  Goldstone modes corresponding to the
symmetry-breaking pattern ${\rm SU}_R(N_f) \times {\rm SU}_L(N_f)
\rightarrow {\rm SU}_V(N_f)$ can be parameterized as
\begin{equation}
  U =  U_R U_L^{-1}
\end{equation}
with $U_R \in {\rm SU}_R(N_f)$ and $U_L \in {\rm SU}_L(N_f)$. The
chiral Lagrangian is constructed by the requirement that it should
have the same invariance properties as the QCD Lagrangian. In
particular, for $m=0$ the Lagrangian should be both Lorentz invariant
and invariant under ${\rm SU}_R(N_f) \times {\rm SU}_L(N_f)$. To
lowest order in the momenta (or derivatives), the kinetic term in the
effective Lagrangian is uniquely given by \cite{Weinberg,GaL}
\begin{equation}
  \label{Lkin}
  {\cal L}_{\rm kin} = \frac {F^2}4 \Tr \partial_\mu U \partial_\mu
  U^{-1}\:.
\end{equation}
The parameter $F$ is the pion decay constant.  The mass term in the
QCD Lagrangian breaks the full flavor symmetry. However, the full
symmetry can be restored if the mass matrix is transformed as well,
\begin{equation}
  M \rightarrow U_L M U_R^{-1}\:. 
\end{equation}
We require that the mass term in the effective Lagrangian satisfies
this extended symmetry. To lowest order in $M$, it is therefore
uniquely given by
\begin{equation}
  \label{Lmass}
  {\cal L}_m = -\frac 12\Sigma\: \Tr (MU + M^\dagger U^{-1})\:.
\end{equation}
For a diagonal mass matrix, the action is therefore
minimized by $U = {\bf 1}$.  The normalization of the mass term is
such that the mass derivative of the partition function according to
Eq.~(\ref{massder}) is equal to the chiral condensate $\Sigma$. The
chiral Lagrangian is valid in the domain
\begin{equation}
  \label{domain1}
  m\ll \Lambda\qquad{\rm and}\qquad p \ll \Lambda\:,
\end{equation}
where $p$ is the momentum and $\Lambda$ is a typical hadronic mass
scale.

We will study the effective Lagrangian at finite volume in a box of
volume $L^4$. Then the smallest nonzero-momentum modes are of the
order
\begin{equation}
  p \sim 1/L\:.
\end{equation}
The fluctuations of the zero-momentum modes, the constant fields, are
not affected by the kinetic term and are limited only by the mass
term.  Comparing Eqs.~(\ref{Lkin}) and (\ref{Lmass}), we see that in a
domain where
\begin{equation}
  \label{domain2}
  \frac {m\Sigma}{F^2} \ll \frac 1{L^2}\:,
\end{equation}
the fluctuations of the zero-momentum modes completely dominate the
fluctuations of the nonzero-momentum modes.  The mass-dependence of
the effective partition function is then given by \cite{GL,Leut92}
\begin{equation}
  Z(m,\theta) = \int_{U\in {\rm SU}(N_f)} DU\: e^{V\Sigma\,
    \Re\Tr Ume^{i\theta/N_f}}\:,
  \label{ZFV}
\end{equation} 
where we have set $M=m{\bf 1}$ and have introduced the
$\theta$-dependence by the substitution $m \rightarrow
m\exp(i\theta/N_f)$ just as for the QCD partition function.  Combining
the conditions (\ref{domain1}) and (\ref{domain2}), we find that the
finite volume partition function is valid in the domain
\cite{GL,Leut92}
\begin{equation} 
  m \ll \frac{1}{\Lambda L^2} \ll \Lambda\:.
  \label{range}
\end{equation}

The partition function for a given topological charge can be extracted
from its $\theta$-dependence.  Since
\begin{equation} 
  Z(m,\theta) = \sum_\nu e^{i\nu\theta} Z_\nu(m)\:,
\end{equation}
we obtain $Z_\nu$ by Fourier inversion. Thus, the finite volume
partition function in the sector of topological charge $\nu$ is
\begin{equation}
  Z_\nu(m) = \frac 1{2\pi}\int d\theta \:e^{-i\nu\theta}Z(m,\theta) 
  =\int_{U\in {\rm U}(N_f)}DU\:{\det}^\nu U
  \:e^{V\Sigma\: \Re\Tr m U}\:,
\label{ZFVnu}
\end{equation}
which coincides with Eq.~(\ref{Znum}).  The partition function
(\ref{ZFVnu}) can be evaluated for arbitrary masses.  The integrals
were first calculated in the context of one-plaquette lattice QCD
models \cite{brower} but later rederived by means of Itzykson-Zuber
integrals~\cite{Russian,Guhr,seneru}.

\subsection{Leutwyler-Smilga Sum Rules}

For masses in the domain of Eq.~(\ref{range}), we can equate the mass
dependence of the QCD partition function and the mass dependence of
the effective partition function.  Equating the coefficients of the
expansion in powers of the mass then gives us the so-called
Leutwyler-Smilga sum rules \cite{Leut92}.  The QCD partition function
can be expanded as
\begin{equation}
  \label{ZQCDnu}
  {Z^{\rm QCD}_\nu(m)} = \frac {\langle 
    m^{|\nu|} \prod_k'(m+i\lambda_k) \rangle}
  {\langle\prod_k' i\lambda_k\rangle}
  = m^{|\nu|}\Bigl( 1 + m^2 \Bigl\langle\sum_{\lambda_k > 0 }\frac
    1{\lambda_k^2}\Bigr\rangle +\ldots\Bigr)\:, 
\end{equation}
where the prime indicates that the product is over nonzero eigenvalues
only.  The effective partition function can be expanded in a power
series in $m$ as well. {}From the U(1) part of the group integral, it
is clear that $Z_\nu^{\rm eff}(m) \propto m^{|\nu|}$. Thus,
\begin{equation}
  \label{Zeffnu}
  Z_\nu^{\rm eff}= m^{|\nu|}(a_0 + a_1 m^2 + \ldots)\:,
\end{equation}
where the coefficients $a_i$ are obtained by calculating the group
integrals. Let us consider two examples. The simplest example is $N_f
=1$, for which
\begin{equation}
  Z_\nu^{\rm eff} = \frac 1{2\pi}
  \int d\theta\: e^{-i\nu \theta} e^{mV \Sigma \cos \theta}
  =I_\nu(mV\Sigma)\:.
\end{equation}
The series expansion of the modified Bessel function is given by
\begin{equation}
  I_\nu(x) = \frac{1}{\nu!}\left (\frac x2\right )^\nu+ \frac 1{(\nu + 1)!} 
  \left (\frac x2\right )^{\nu+2} + \ldots 
\end{equation}
for $\nu\ge0$.  We also have $I_{-\nu}(x)=I_\nu(x)$.  Matching the
normalizations of Eqs.~(\ref{ZQCDnu}) and (\ref{Zeffnu}) and equating
the coefficients of the $O(m^2)$ terms, we find
\begin{equation}
   \left\langle\frac 1{V^2} \sum_{\lambda_k > 0} \frac
  1{\lambda_k^2}\right\rangle= \frac{\Sigma^2}{4(|\nu|+1)}\:.
\end{equation} 
A second example is the case of $\nu = 0$ and arbitrary $N_f$. In this
case, we use the group integral
\begin{equation}
  \int DU \:U_{ij} U^\dagger_{kl} = \frac 1{N_f} \delta_{jk} \delta_{il}
\end{equation}
to derive the result
\begin{equation}
  \left\langle\frac 1{V^2} \sum_{\lambda_k > 0}\frac
    1{\lambda_k^2}\right\rangle = \frac{\Sigma^2}{4N_f}\:.
\end{equation} 
Since asymptotically for large $\lambda$ the spectral density is
proportional to $V\lambda^3$, the sum over the eigenvalues has to be
regularized. As discussed above in Sec.~\ref{diracspectrum}, a
finite result is obtained by taking the thermodynamic limit before
removing the cutoff.

What do we learn from these sum rules? Since the total number of
eigenvalues is of order $V$, the obvious interpretation is that the
smallest eigenvalues are of order $1/V$. Indeed, this is in agreement
with the Banks-Casher formula. What is more important,
however, is that in the derivation of the sum rules we have
relied only on the chiral symmetry of the partition function and its
spontaneous breaking by the formation of a chiral condensate.
Therefore any theory with the same pattern of chiral symmetry breaking
as QCD should obey the same spectral sum rules. In particular, the
eigenvalues distributed according to the chRMT
introduced in the previous section should obey the same sum rules.  In
fact, all Leutwyler-Smilga sum rules can be derived systematically
from the joint eigenvalue chRMT probability distribution in
Eq.~(\ref{zeig}) using the known Selberg integrals \cite{sell}. In
addition to these sum rules, it is possible to derive massive sum rules
\cite{lsmass}.  A relation between sum rules and partially quenched
effective theories was considered in Ref.~\cite{zyab}.  Spectral sum rules
for Ginsparg-Wilson fermions on the lattice were considered in
Ref.~\cite{Farc99c}.  Inverse moments of the small Dirac eigenvalues
in generalized chiral perturbation theory were derived in
Ref.~\cite{Desc99}. 

We have seen that chRMT allows us to
calculate the correlations of the smallest eigenvalues of the Dirac
operator on the scale of the average level spacing. This raises the
question of whether it is possible to derive the microscopic properties
of the eigenvalues from the low-energy effective partition function.
The answer is no.  The spectral density cannot be derived from the
mass dependence of the chiral condensate,
\begin{equation}
  \Sigma(m) =\left \langle\frac 1V \sum_k \frac
    1{i\lambda_k+m}\right\rangle,
\end{equation}
because the average contains the fermion determinant with the same
mass. As we show in the next section, this problem can be
circumvented by the introduction of a spectral mass.

\subsection{Spectral Mass and Partially Quenched Partition Function}
\label{SpectralMass}

The Dirac spectrum can be obtained from the resolvent
\begin{equation}
  \Sigma(z) =\left\langle \frac 1V\: \Tr \frac 1{{\cal D}+z} \right
  \rangle\:, 
\end{equation}
were the spectral mass $z$ is an independent complex variable that
does not occur in the average. With purely imaginary eigenvalues the
spectral density is given by the discontinuity of the resolvent across
the imaginary axis,
\begin{equation}
  \frac {\rho(\lambda)}V = \lim_{\epsilon\to0}\frac 1{2\pi}
  [ \Sigma(i\lambda +\epsilon) - \Sigma(i\lambda -\epsilon)]\:.
  \label{discon}
\end{equation}
This follows immediately upon writing the trace as a sum over the
eigenvalues of ${\cal D}$.  The generating function for the resolvent
can be written down easily \cite{Vminn},
\begin{equation}
  Z^{\rm sp} = \left \langle \frac {{\det}^{N_f}({\cal D}+m)
      \det({\cal D}+z)}{\det({\cal D}+z')} \right \rangle_{\rm YM}\:,
  \label{Zsp}
\end{equation}
where we have explicitly displayed the fermion determinant in the
measure.  The resolvent is then given by
\begin{equation}
  \Sigma(z) = \left. \frac 1V \partial_z Z^{\rm sp}(z,z')\right|_{z'=z}.
  \label{resol}
\end{equation}
In QCD, this method was first introduced in order to derive the
quenched strong coupling expansion \cite{Morel}. In nuclear physics
and condensed matter physics, this method is known as the
supersymmetric method or Efetov method for quenched disorder
\cite{Brezin,Efetov,VWZ}.

The determinants in the numerator can be written as fermionic
integrals where\-as the determinant in the denominator can be written
as a bosonic integral. The partition function (\ref{Zsp}) is thus
invariant under flavor symmetries that mix commuting and
anticommuting degrees of freedom. More precisely, for $m=z=z'=0$ the
partition function is invariant under the supergroup ${\rm
  Gl}_R(N_f+1|1) \times {\rm Gl}_L(N_f+1|1)$. In order to obtain a
consistent effective theory, it is essential to extend the unitary
symmetry to the general linear group Gl. We expect, as for the
QCD vacuum, that chiral symmetry is spontaneously broken to
the diagonal subgroup ${\rm Gl}_{\rm V}(N_f+1|1)$.  The mass term explicitly
breaks the full symmetry to this subgroup as well.
 
The effective partition function can be derived in the same way as the
regular chiral Lagrangian with one complication. To obtain finite
integrals, we must make sure that the integration manifold is
Riemannian. This is why we have extended the flavor
symmetry to the full general linear group.  The integration manifold
is then given by the maximum Riemannian submanifold of ${\rm
  Gl}_{\rm A}(N_f+1|1)$. In plain language, this means that we compensate the
extra minus sign in the supertrace by complexifying the group
parameters.  The generating function is thus given by \cite{OTV}
\begin{equation}
  \label{zspect}
  Z^{\rm pq}=\hspace*{-17pt}\int\limits_{U\in {\hat {\rm Gl}}(N_f+1|1)} 
  \hspace*{-17pt} DU e^{\int d^4x \left[-\frac
      {F^2}4 \Str \partial_\mu U \partial_\mu U^{-1} -
      \frac{F^2m_0^2}{12}\left(\frac{\sqrt 2 \Phi_0}F  
        - \theta \right )^2 + \frac{ \Sigma}2\:
      \Str (MU + M^\dagger U^{-1})\right]}\:, \nonumber \\
\end{equation}
where $i\sqrt 2\Phi_0/F = {\rm Str} \log U $.  This partition function
is also known as the partially quenched effective partition function.
It has been used to obtain a better understanding of quenched lattice
QCD results \cite{Maarten,Sharpe}.  The hat on Gl denotes the maximum
Riemannian submanifold of ${\rm Gl}(N_f+1|1)$, and Str stands for the
supertrace.  An example of an explicit parameterization of $U$ will
be given in the next subsection.  The parameter $m_0^2$ is
proportional to the topological susceptibility and results in a mass
for the singlet channel that does not vanish in the chiral limit.

The perturbative formulation of this partition function was first
given by Bernard and Golterman \cite{Maarten}, and a perturbative
calculation of the dependence of the condensate on the spectral mass
was first performed in Ref.~\cite{Leung}. In that case, it is
acceptable to ignore
the convergence properties of the effective partition function and
integrate over the unitary supergroup instead.

An alternative generating function would be obtained by replacing the
ratio of 
the two determinants by ${\det}^n({\cal D}+z)$ and putting $n
\rightarrow 0$ after having calculated the resolvent.  This procedure,
known as the replica trick \cite{edwards}, successfully
reproduces the asymptotic expansions of the spectral correlation
functions \cite{asympt,pourep}.  There have been recent claims that it
is possible to derive truly nonperturbative results by means of the
replica trick \cite{kamenev,yurk,kanrep}.  We do not believe that this
can be done in general \cite{critique,zirnrep,dalmazi}.  As mentioned,
because of the compact/noncompact structure of the final answer, the
supersymmetric formulation is the natural approach to this problem.

\subsection{Domains of the Partially Quenched Effective Theory}
\label{sec:domains}

The Goldstone fields can be written as
\begin{equation}
  \label{Ufields}
  U = e^{i\sqrt 2\Pi/F}\:.
\end{equation}
To second order in the fields, the effective Lagrangian in momentum
space is given by
\begin{equation}
  {\cal L} = \frac 1V \sum_a\sum_k  (k^2 + M_a^2) \Pi_a^2(k)\:,
  \label{L2}
\end{equation}
where the sum is over the momenta in a box of length $L$ (including
the zero-momentum state) and the masses of the Goldstone bosons are
denoted by $M_a$. The sum over $a$ is over the different Goldstone
modes.  In addition to the usual Goldstone modes with mass
\begin{equation}
  \label{Gmass1}
  M_{mm}^2 = \frac{2m\Sigma}{F^2}\:,
\end{equation}
there are both fermionic and bosonic Goldstone bosons with mass
\begin{equation}
  M_{mz}^2 = \frac{(m+z)\Sigma}{F^2}
\end{equation}
and fermionic and bosonic Goldstone bosons with mass
\begin{equation}
  \label{Gmass3}
  M_{zz}^2 = \frac{2z\Sigma}{F^2}\:.
\end{equation}
In the Lagrangian (\ref{L2}), one can
distinguish the zero-momentum modes from the nonzero-momentum modes.
The magnitude of the fluctuations of the nonzero-momentum modes is of
order $1/(k^2+M^2_a)$, whereas the magnitude of the fluctuations of
the zero-momentum modes is of order $1/M_a^2$. Since the smallest
nonzero momenta are of order $1/L$, the fluctuations of the
zero-momentum modes are dominant if 
\begin{equation}
  \label{compton}
  M_a^2 \ll \frac 1{L^2}\:.
\end{equation}
This means that the Compton wavelength of the ``pion'' is much larger
than the size of the box.  For nonzero quark masses of order $O(L^0)$,
the inequality (\ref{compton}) is never satisfied. However, the
spectral mass $z$ is a free parameter, and the inequality can be
rewritten as \cite{Verb96,Trento}
\begin{equation}
  z \ll E_c\equiv\frac {F^2}{\Sigma L^2}\:.
  \label{spectralrange}
\end{equation}
We sometimes refer to the quantity $E_c$ as the Thouless energy 
for reasons that will become clear in Sec.~\ref{sec:anderson}.  In the
domain of Eq.~(\ref{spectralrange}), the dominant contributions to the
resolvent are from the zero-momentum modes.  Thus, the partially
quenched effective partition function (\ref{zspect}) can be reduced to
the partition function in the zero-momentum sector \cite{OTV,Damg99},
\begin{equation}
  \label{static}
  Z= \int_{U\in {\hat{\rm Gl}}(N_f+1|1)} DU e^{\frac12 
    V\Sigma\:\Str(MU + M^\dagger U^{-1})}\:.
\end{equation}
The microscopic spectral density and the spectral correlations
computed from Eq.~(\ref{static}) are identical to those obtained in chRMT
\cite{Verb96,Trento}.  Let us show this by an explicit calculation of
the dependence of the chiral condensate on the spectral mass $z$. To
cover the complete range (\ref{spectralrange}), the integral over $U$
has to be done nonperturbatively.  This is a straightforward
superintegral that can be evaluated using standard methods. The
calculation of the integration measure requires an explicit
parameterization of the integration manifold. For example, in the
quenched limit, $N_f = 0$, a possible choice is
\begin{equation}
  U = \mat e^{i\phi} & \alpha \\ \beta & e^s \emat
\end{equation} 
with $\alpha $ and $\beta$ Grassmann variables, $\phi \in [0, 2\pi]$,
and $s \in (-\infty, \infty)$. A characteristic feature of the
integration manifold is that it consists of a compact and a noncompact
component.  The valence-quark mass dependence of this partition
function and its generalization to arbitrary topological charge
coincide with the valence quark mass dependence of the chUE partition
function.  We merely quote the final result for the resolvent for $N_f$
massless flavors in the sector of topological charge $\nu$
\cite{Verb96},
\begin{equation}
  \label{valk}
  \frac{\Sigma(u)}{\Sigma} = u\left[I_{a}(u)K_{a}(u) 
    +I_{a+1}(u)K_{a-1}(u)\right]+\frac{|\nu|}{u}\:,
\end{equation}
where $ a= N_f+|\nu|$ and $u= zV\Sigma$.  The compact/noncompact
symmetries are reflected in the appearance of the $I_a/K_a$-Bessel
functions and are thus a natural ingredient of the underlying
integration manifold. This result was first obtained from chRMT by
integrating the microscopic spectral density in Eq.~(\ref{micro2}) as
follows,
\begin{equation}
 \frac{\Sigma(u)}{\Sigma} = \int_0^\infty d\zeta 
 \frac{2u}{\zeta^2+u^2}\rho_s(\zeta)\:. 
\end{equation}
Alternatively, the microscopic spectral density can be obtained by
taking the discontinuity of $\Sigma(u)$ according to
Eq.~(\ref{discon}).  By integrating back, we find that the mass
dependence of the zero-momentum partially quenched partition function
coincides with that of the chRMT partition function for $\beta=2$.

\begin{figure}[t]
  \begin{center}
    \epsfig{figure=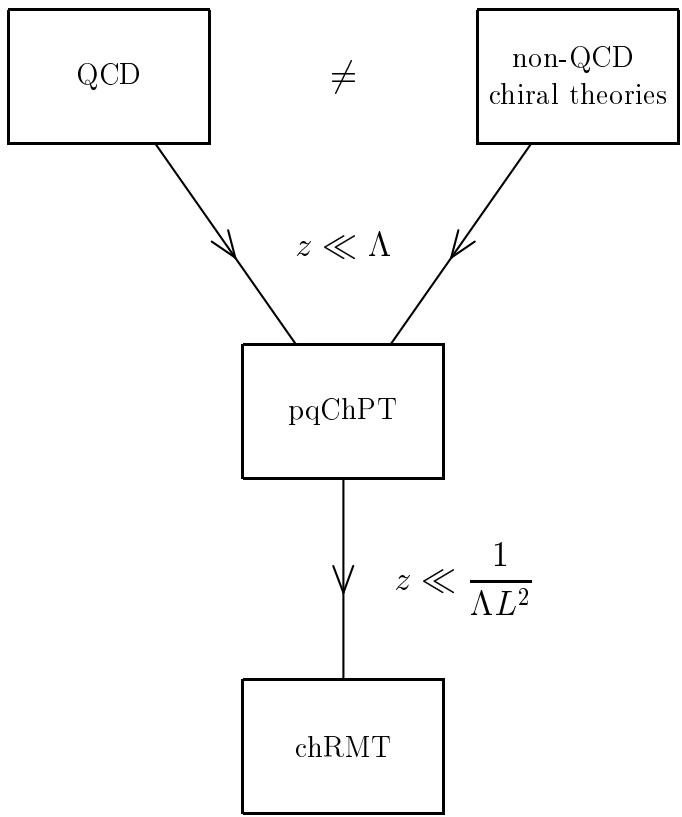,width=90mm}
    \caption{Diagram illustrating the different domains in the QCD
      Dirac spectrum.  Here,  pqChPT denotes partially quenched chiral
      perturbation theory, and chRMT stands for chiral random matrix
      theory.}  
    \label{fig:domains}
  \end{center}
\end{figure}

Figure~\ref{fig:domains} presents a schematic picture of the
different domains in the Dirac spectrum. There are other non-QCD theories that
can be reduced to the same partially quenched partition function.
Probably the best known example is the instanton liquid model of QCD
\cite{instantonl}. Another example, more closely related to
disordered condensed matter systems, is the random-flux model
\cite{simons-altland}, which is in essence quenched lattice QCD with
Kogut-Susskind fermions but without the phase factors due to the
$\gamma$-matrices.  Related examples are so-called two-sublattice
models with disorder \cite{Gade}, and disordered lattice models with the
chiral and flavor symmetries of QCD \cite{Takahashi,Guhr-Wilke-Weidi}.

In the range
\begin{equation}
  E_c \ll z \ll \Lambda\:,
  \label{dom1}
\end{equation}
the effective action (\ref{zspect}) is still valid but chiral random
matrix theory no longer applies. In this range, the $U$ fields can be
expanded to second order in the pion fields, and the spectral mass
dependence of the chiral condensate can be obtained to one loop order.
The spectral density follows by taking the discontinuity of
$\Sigma(z)$ across the imaginary axis. This calculation can be
performed both for the partition function of Eq.~(\ref{zspect}) and the
effective partition functions for $\beta =1$ and $\beta =4$. The
result for the slope of the spectral density at zero, valid for
$N_f\ge2$, reads \cite{Smilstern,Toub99}
\begin{equation}
  \frac{\rho'(0)}{\rho(0)} =
  \frac {(N_f-2)(N_f+\beta)}
  {16\pi \beta N_f }\frac{\Sigma }{F^4}\:.
\end{equation}
The result for $\beta =2$ was first derived by Smilga and Stern
\cite{Smilstern} from the scalar susceptibility in standard chiral
perturbation theory.  We emphasize that this is an exact result for
the QCD Dirac spectrum that is valid in the thermodynamic limit.  The
vanishing of the slope for $N_f =2$ has been confirmed by instanton
liquid simulations \cite{Acta}.

The two-point correlation function can also be obtained from a
supersymmetric generating function with, in this case, two different
spectral masses and two bosonic superpartners \cite{OTV}.  Again the
zero-momentum contribution coincides with the chUE result. Let us
consider the perturbative expansion of the two-point correlation
function.  Generalizing Eq.~(\ref{discon}), we obtain a relation
between the two-point level correlation function and the pion
susceptibility \cite{OTV,carter},
\begin{equation}
  \langle \rho(\lambda) \rho(\lambda') \rangle^C =
  \frac 1{4\pi^2}\frac {\Sigma^2}{F^4}\left . {\rm Disc}
  \right |_{z = i\lambda, z'=i\lambda'}\sum_q\frac 1{(q^2 +M^2)^2},
\end{equation}
where the meson mass is given by
$M^2=(\sqrt{z^2}+\sqrt{z'^2})\Sigma/F^2$ and the 
superscript $C$ denotes the connected part of the two-point function.
By taking the discontinuities of the zero-momentum contribution, we
find the two-point correlation function \cite{OTV}
\begin{equation}
  \langle \rho(\lambda) \rho(\lambda') \rangle^C \sim
  -\frac 1{2\pi^2} \left [\frac 1{(\lambda-\lambda')^2}
  + \frac 1{(\lambda+\lambda')^2}\right ]\:,
\end{equation}
which is the correct asymptotic result for the chUE.

\subsection{QCD in Three Euclidean Dimensions}

QCD in three Euclidean dimensions can be analyzed in much the same way as
discussed above. The main difference is the absence of the ${\rm
  U_A}(1)$ symmetry and the absence of instantons. In terms of the
Dirac spectrum, the eigenvalues do not occur in pairs $\pm i\lambda$,
and strictly zero eigenvalues are absent. The Dirac spectrum of this
theory on the microscopic scale can be analyzed along the same lines
as the four-dimensional theory.  The microscopic
spectral density and the Leutwyler-Smilga sum rules have been derived
\cite{VZ3,hilmoine}, the low-energy effective theory has been identified
\cite{VZ3,magnea}, the mass-dependent microscopic spectral density has
been found \cite{massuni,nagnish}, universality has been studied
\cite{Akem97,christian}, and lattice QCD studies have confirmed the
theoretical analysis \cite{Damg98b}.

\section{UNIVERSAL PROPERTIES OF THE LATTICE QCD DIRAC SPECTRUM}
\label{sec3}

\subsection{Unfolding}
\label{unfolding}

In order to compare eigenvalues of any physical system with RMT
results, it is necessary
first to ``unfold'' the empirical spectrum (see e.g.\
Ref.~\cite{Bohi84a}).  The spacing $\Delta$ of the eigenvalues is
related to the spectral density by
\begin{equation}
  \Delta(E)=1/\rho(E)\:.
\end{equation}
Unfolding is a local rescaling of the energy scale so that the mean
level spacing of the unfolded eigenvalues is equal to unity throughout
the spectrum.  This is achieved by splitting the spectral density in
an average density, $\bar \rho(E)$, and a fluctuating piece,
\begin{equation}
  \rho(E) = \bar \rho(E) + \rho_{\rm fl}(E)\:.
\end{equation}
The average spectral density can be obtained in different ways. In
some cases, it can be computed analytically using semiclassical
arguments.  However, in most cases $\bar \rho(E)$ is obtained by
averaging over many level spacings. There are two essentially
different procedures to do this, spectral averaging and ensemble
averaging.  Spectral averaging is appropriate if only one or a few spectra
are available. The average level spacing at $E$ is then obtained by
averaging over many level spacings around $E$.  Ensemble
averaging is appropriate if there is a large ensemble of spectra all
drawn from 
the same statistical distribution. In this case, the average spacing
can be obtained by averaging the spacing at $E$ over each member of
the ensemble.  In both cases, the separation of the spectral density
into an average and a fluctuating piece requires a separation of
scales that is typically achieved only in the thermodynamic limit or
in the semiclassical limit.  The unfolded spectrum, $\{x_n\}$, is then
obtained from the average spectral density (equal to the inverse
average level spacing) by
\begin{equation}
  x_n=\int_{-\infty}^{E_n}\bar\rho(\lambda) d\lambda\:,
\end{equation}
where $\{E_n\}$ is the original sequence of eigenvalues.  The
difference between the two procedures is that spectral averaging yields
$\bar\rho(E)$ separately for each spectrum whereas ensemble averaging
yields a single $\bar\rho(E)$ for all spectra.  In both cases, the
average spectral density of the unfolded spectrum becomes
$\bar\rho(x)=1$, so that the sequence $\{x_n\}$ has an average level
spacing equal to unity.  The correlations of the levels are always
calculated for the unfolded spectrum.

The two procedures to calculate the average spectral density
do not necessarily give the same spectral correlations at long
distances. The equivalence of the two is known as spectral
ergodicity.  This property has been shown analytically for several
random matrix ensembles \cite{Pandey,Weide-recent}.  In general,
ensemble averaging results in stronger level fluctuations than does
spectral averaging \cite{French-connection,Guhr99,Ande99}.  In practice
a mixture of both methods is often useful. For example, one may calculate
the average spectral density by ensemble averaging but, in order to
get better statistics, calculate the correlation functions by both
ensemble averaging and spectral averaging.  Spectral averaging requires
that the statistical properties of the eigenvalues be stationary over the
spectrum, which is not the case for many systems and has to be checked
each time.

\subsection{Lattice Tests of Chiral Random Matrix Theory}
\label{LatticeTests}

We have presented a wealth of analytical evidence supporting
the statement that the local spectral correlations of the Dirac
operator are described by universal functions.  Although the Dirac
spectrum is not directly observable in experiments, we can compare the
predictions of chiral RMT to numerical data obtained by lattice gauge
simulations.  This is the subject of this section.

This review does not discuss the global spectral density of the
QCD Dirac operator.  Most lattice results have large finite-size
artifacts.  Instanton simulations suggest, contrary to random
matrix results, that the fermion determinant has a significant effect on
the global Dirac spectrum \cite{Verb94c,Acta,Osbo98a,Sharan}.

\subsubsection{BRIEF INTRODUCTION TO LATTICE QCD}

QCD is a renormalizable quantum field theory that must be regularized.
This can be done by formulating the theory on a discrete lattice with
lattice spacing $a$ \cite{Wils74}.  The largest momentum is then
$\pi/a$.  The quark fields live on the sites and the gauge fields on
the links of the lattice.  If the lattice is finite, one can simulate
the theory on a computer.  This can be done efficiently only in
Euclidean space, where the gluonic weight function is $\exp(-S_{\rm
  YM})$ with $S_{\rm YM}$ real.  The discretized form of the
Yang-Mills action $S_{\rm YM}$ is the Wilson action $S_{\rm W}$.  The
full weight function of QCD also contains the fermion determinants,
which can be expressed in terms of the gauge fields.  Observables are
computed by generating gauge field configurations in a Monte Carlo
update procedure and averaging an observable over many configurations.
Since the inclusion of the fermion determinants is very
time consuming, it is common to use only the gluonic part of the
weight function in the Monte Carlo updates.  This is called the
quenched approximation, which corresponds to the limit $N_f=0$ or,
equivalently, to the limit of infinitely heavy sea quarks.  To make
contact with continuum physics, the results of lattice simulations
must be extrapolated to infinite lattice size (the thermodynamic limit)
and to zero lattice spacing (the continuum limit).

Lattice simulations are the main source of nonperturbative information
about QCD.  Unfortunately, the naive discretization of fermions on a
lattice leads to the so-called doubling problem: The quark propagator
has poles at each corner of the Brillouin zone, which gives rise to a
total of $2^d$ species in $d$ dimensions.  The unwanted $2^d-1$
species can be eliminated by adding to the Dirac operator an
additional term, the Wilson term, which removes the doublers in the
continuum limit.  However, this term breaks chiral symmetry
explicitly.  Another possibility is the use of staggered (or
Kogut-Susskind) fermions where one has only one spinor component per
lattice site.  This maintains a residual chiral symmetry but only
partially reduces the number of species to $2^{d/2}$ in the $a\to0$
limit.  A no-go theorem by Nielsen and Ninomiya \cite{Niel81}
states that it is not possible simultaneously to solve
the doubling problem and have exact chiral symmetry on the lattice
with a local action. 

Fortunately, there is a way around this theorem: the remnant chiral
symmetry condition of Ginsparg and Wilson \cite{Gins82},
\begin{equation}
  \label{GWC}
  {\cal D}\gamma_5+\gamma_5{\cal D}=2a{\cal D}\gamma_5R{\cal D}\:,
\end{equation}
where $R$ is a spatially local operator that is trivial in Dirac
space.  In contrast to $\{{\cal D},\gamma_5\}=0$, chiral symmetry is
not exact but is recovered only in the continuum limit $a\to0$.  Thus,
with Dirac operators satisfying Eq.~(\ref{GWC}), it is possible to get rid of
the doublers even at finite lattice spacing without violating the
Nielsen-Ninomiya theorem.  Recently, several solutions of
Eq.~(\ref{GWC}) have been found in the overlap formalism
\cite{Neub98}, in the domain-wall formulation in five dimensions
\cite{Kapl93}, and in the perfect action approach \cite{Hase98}.

As mentioned in Sec.~\ref{anti-unitary}, there may be cases in which the
anti-unitary symmetries of the various lattice discretizations of the
Dirac operator differ from those of the continuum operator.  In
particular, this is true for staggered fermions.  In SU(2) color,
staggered fermions are in the symmetry class of the chSE, whereas
continuum fermions are in the chOE symmetry class.  Staggered fermions
in the adjoint representation of the gauge group (for any $N_c$) have
the symmetries of the chOE, whereas in the continuum limit the
symmetries are those of the chSE.  The Wilson Dirac
operator ${\cal D}_{\rm W}$ does not anticommute with $\gamma_5$ and
is therefore not described by any of the chiral ensembles.  However,
the Hermitian operator $\gamma_5{\cal D}_{\rm W}$ has the same
anti-unitary symmetries as the continuum Dirac operator and is
described by the corresponding nonchiral ensembles (see also
Ref.~\cite{Hehl}).  Finally, Dirac operators obeying the
Ginsparg-Wilson condition, Eq.~(\ref{GWC}), have the same anti-unitary
symmetries as the continuum Dirac operator, but the eigenvalues are
located on the complex unit circle \cite{Splittorff}.

\subsubsection{BULK CORRELATIONS}

To measure the bulk spectral correlations, one needs all eigenvalues
of the Dirac operator, which is represented on the lattice by a finite
sparse matrix whose dimension is proportional to the lattice volume
and to $N_c$.  The eigenvalues of such a matrix can be obtained using
special algorithms, e.g.\ by the Cullum-Willoughby version of the
Lanczos algorithm \cite{Cull81}.  On the lattice, this was first done
by Kalkreuter \cite{Kalk95}.  The numerical effort of this method
scales with the square of the matrix dimension.  For example, in SU(3)
on a $10^4$ lattice, the Dirac matrix has 15,000 distinct positive
eigenvalues, which can be computed on a typical workstation in about
40 minutes.  There are exact sum rules for the sum of the squares of
all Dirac eigenvalues, which can be used to check the numerical
accuracy.  Because the number of eigenvalues per configuration is very
large and because the ensemble average can be replaced by a spectral
average under the assumption of spectral ergodicity and stationarity,
one needs only a few configurations to construct the bulk spectral
correlations with great accuracy.

\begin{figure}
  \begin{center}
    \epsfig{figure=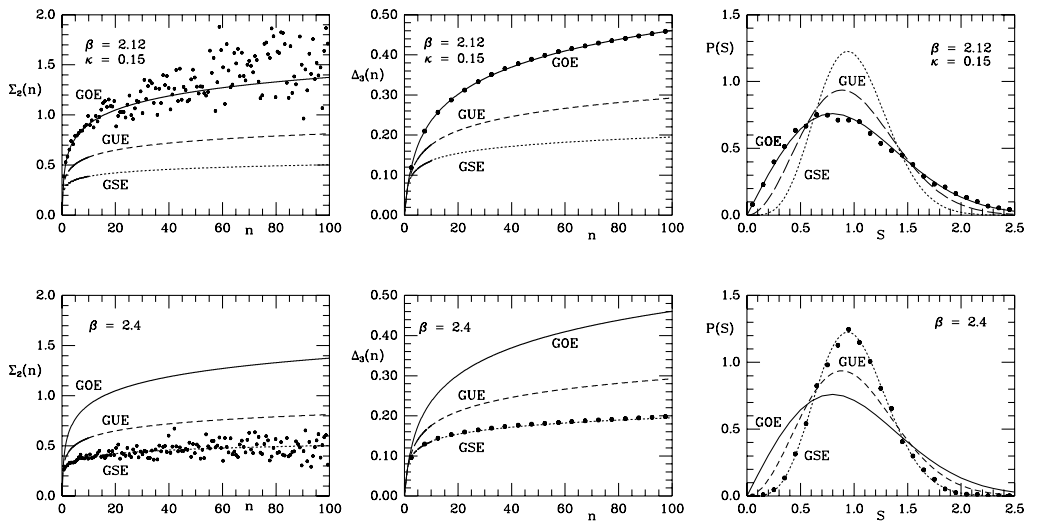,angle=270,width=\textwidth}
    \caption{Number variance $\Sigma^2(n)$, spectral rigidity
      $\Delta_3(n)$, and nearest-neighbor spacing distribution $P(s)$
      of the lattice Dirac operator with gauge group SU(2).  {\it Top:}
      Wilson fermions, $V=8^3\times12$, $N_f=2$.  {\it Bottom:}
      staggered fermions, $V=12^4$, $N_f=4$, $ma=0.05$.  (From
      Ref.~\protect\cite{Verb97a}. Note that in
      Ref.~\protect\cite{Verb97a}, the figure of $P(s)$ for Wilson
      fermions shows data points for staggered fermions. This has been
      corrected in the present figure.)}
    \label{bulk}
  \end{center}
\end{figure}

The bulk correlations have been measured in a number of lattice
studies for all three symmetry classes
\cite{Hala95a,Verb97a,Pull98,Guhr99,Berg99,Edwa99b} and in instanton
liquid simulations \cite{Osbo98a} for $\beta =2$.  As mentioned
above, they are insensitive to the number of flavors and the
topological charge.  In all cases, excellent agreement with the
predictions of the appropriate random matrix ensemble was obtained
(see e.g.\ Fig.~\ref{bulk}).  
The agreement is perfect not only in
the strong-coupling regime but also at weak coupling.  In fact, the
bulk spectral correlations are given by RMT even in the deconfinement
phase \cite{Pull98,Berg99}, indicating that the gauge fields retain a
sufficient degree of randomness in this region of the phase diagram.

Spectral ergodicity was investigated in Ref.~\cite{Guhr99}, and the
equivalent of a Thouless energy was found for ensemble averaging,
whereas spectral averaging resulted in complete agreement with RMT
correlations over distances as long as several hundred average level
spacings.

\subsubsection{MICROSCOPIC CORRELATIONS}

Because only the lowest eigenvalues contribute to the microscopic
spectral correlations, a large number of statistically
independent spectra are necessary.  In contrast to
the bulk correlations, the microscopic correlations are sensitive to
the number of flavors and the topological charge.

The RMT predictions do not contain an energy scale.  In order to make
comparisons with lattice data, one needs to determine the energy scale
$1/V\Sigma$ to be used in the RMT expressions, see e.g.\
Eq.~(\ref{micro}).  This can be done by extracting $\rho(0)$ from the
data and applying the Banks-Casher relation, Eq.~(\ref{BC}).  Because
this procedure makes no reference to RMT, the comparisons of lattice
data with RMT are parameter-free.

\begin{figure}
  \begin{center}
    \epsfig{figure=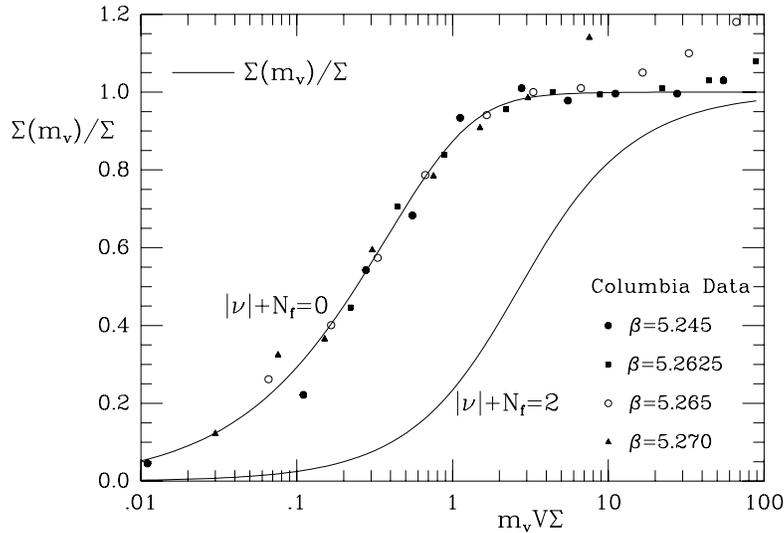,width=70mm,angle=270}
    \caption{Valence quark mass dependence of the chiral condensate
      $\Sigma(m_v)$ plotted as $\Sigma(m_v)/\Sigma$ versus $m_v
      V\Sigma$.  The dots and squares represent lattice results by the
      Columbia group \cite{Chan95} for the values of $\beta$ indicated
      in the figure.  The solid lines are chRMT results.  (From
      Ref.~\cite{Verb96}.)}
    \label{fig5}
  \end{center}
\end{figure}

The first numerical results for the microscopic spectral density
$\rho_s$ were obtained using instanton liquid configurations for
$N_c=2,3$ and $N_f=0,1,2$ \cite{Verb94c}, and the expected agreement
with the corresponding random matrix predictions was found.  The
microscopic spectral density was first observed on the lattice via the
dependence of the chiral condensate on a valence quark mass $m_v$, as
studied by the Columbia group \cite{Chan95}.  Figure~\ref{fig5} 
shows $\Sigma(m_v)/\Sigma$ as a function of $m_vV\Sigma$ for different
values of the coupling constant \cite{Verb96}.  The
data for different coupling strengths in the broken phase fall on a
single curve and agree with the chRMT result for
$\Sigma(m_v)$ for $N_f = |\nu |= 0$, see Eq.~(\ref{valk}).  This
figure also shows a deviation from the chRMT predictions at a
scale that is of the order of the Thouless energy given in
Eq.~(\ref{spectralrange}).
The spectral mass dependence of the chiral condensate has recently
been studied for different symmetry classes and different types of
fermions, and a similar quality of agreement has been found
\cite{Damg99b,Hernandezmv}.

\begin{figure}
  \begin{center}
    \epsfig{figure=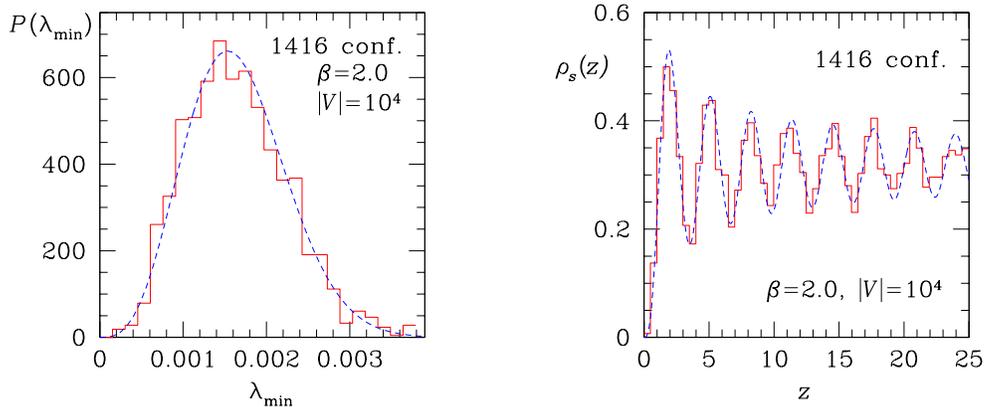,width=130mm}
    \caption{Distribution of the smallest eigenvalue {\it (left)} and
      microscopic spectral density {\it (right)} of the staggered Dirac
      operator in quenched SU(2).  The dashed lines are the predictions
      of the chSE for $N_f=0$ and $\nu=0$.  (From
      Ref.~\cite{Berb98a}.)} 
    \label{figmicro}
  \end{center}
\end{figure}

The microscopic spectral correlations have also been investigated in
great detail on the lattice
\cite{Berb98a,Ma98,Damg98b,Damg99a,Gock98,Edwa99a,Damg99b}, and the
random matrix predictions were confirmed with very high accuracy for
all three symmetry classes.  A typical example is shown in
Fig.~\ref{figmicro}.

Although most of these simulations have been performed in the quenched
approximation, it is also possible to include dynamical fermions.
Deviations from the $N_f=0$ results are observed only if the sea
quarks are very light, with masses on the scale $1/V\Sigma$.
Otherwise, the mass term in Eq.~(\ref{eq1}) dominates the small
eigenvalues in the factors of $(i\lambda_n+m_f)$.  Analytical results
for the microscopic spectral correlations in the presence of sea
quarks with mass $\sim1/V\Sigma$ were obtained in
Refs.~\cite{Jurk96,Damg97,Wilk98,Nishim,akemann-kanzieper}.  Lattice
results in this regime agree very well with the corresponding RMT
predictions \cite{Berb98b}.  In the Schwinger model, it is numerically
feasible to consider massless sea quarks, and the lattice data are
again well described by RMT \cite{Farc98}.

Recall that the energy scale for the small eigenvalues is $1/V\Sigma$.
Given the agreement of the microscopic spectral quantities with RMT,
this means that RMT can be used to determine
the infinite-volume chiral condensate $\Sigma$ \cite{tilo-extra}.
This is most easily done by fitting the numerically determined
distribution of the smallest eigenvalue to the RMT result.  Since the
lattice volume is known, this immediately yields $\Sigma$.

\subsubsection{TOPOLOGY}

The random matrix predictions for the microscopic spectral
correlations depend on the number of zero modes of the Dirac operator
and, thus, on the topological charge $\nu$ of the gauge field
configurations.  Thus, one should sort the configurations according to
their values of $\nu$ and make the comparison with RMT separately in
sectors of fixed topological charge.

The results presented in the previous section were obtained for the
staggered Dirac operator. If one is interested in topological
properties, this operator has problems. The zero modes that one would
obtain in the continuum limit are shifted at finite lattice spacing
$a$ by an amount proportional to $a^2$ \cite{Smit87}.  For reasonable
simulation parameters, this amount is larger than the level spacing
near zero so that the would-be zero modes are completely mixed with
the nonzero modes.  Thus, one should expect all spectra of the
staggered Dirac operator, even if computed from gauge field
configurations with nonzero topological charge, to be described by the
RMT results for $\nu=0$.  This was indeed observed in the early data
\cite{Verb96,Berb98a} and more recently confirmed in a detailed study
(though only in strong coupling) \cite{Damg99c}.  As the continuum
limit is approached, the would-be zero modes should move toward
$\lambda=0$ and eventually separate completely from the nonzero modes.
This effect has recently been observed in the Schwinger model in two
dimensions \cite{Farc99a}.

\begin{figure}
  \begin{center}
    \epsfig{figure=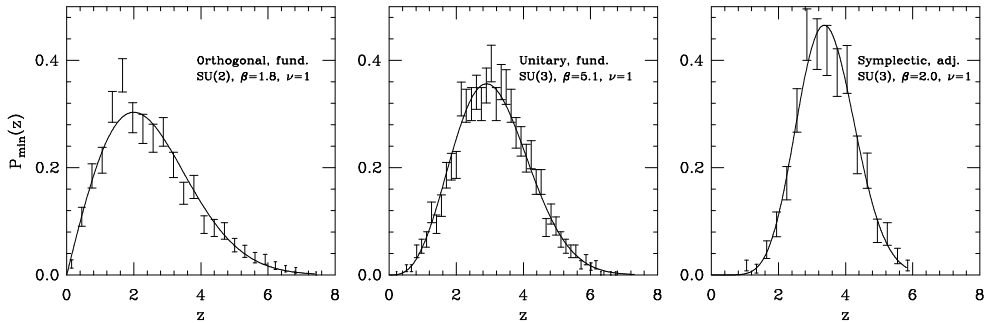,width=130mm}
    \caption{Distribution of the smallest Dirac eigenvalue in the
      $\nu=1$ sector for all three symmetry classes.  The data were
      obtained using the overlap Dirac operator on a $4^4$ lattice.
      Solid lines represent the corresponding RMT results.  (From
      Ref.~\cite{Edwa99b}.)}
    \label{top}
  \end{center}
\end{figure}

Fortunately, Dirac operators satisfying the Ginsparg-Wilson condition
(\ref{GWC}) do not have this problem.  The overlap
operator \cite{Neub98} has exact zero
modes even at finite lattice spacing, and the microscopic spectral
correlations in sectors of fixed $\nu$ are in perfect agreement with
the corresponding random matrix predictions
\cite{Farc98,Edwa99b,Damg99b,Hernandezmv,Tilos}.   Figure~\ref{top}
shows an example.

\subsubsection{CHIRAL PHASE TRANSITION}
\label{sec:chphtr}

As mentioned in Sec.~\ref{UniversalityProofs}, the microscopic
correlations are given by chiral RMT only if $\rho(0)/V>0$, i.e.\ if
chiral symmetry is spontaneously broken.  Chiral symmetry is restored
above a critical temperature $T_c$.  The way in which $\rho(0)/V$
approaches zero determines the order of the chiral phase transition
and, if it is second-order, the associated critical exponents.
Lattice investigations of these questions are very difficult because at
finite volume there are no sharp phase transitions.  Simulations are
plagued by 
finite-size effects, critical slowing down, and other problems.  It
is therefore of great interest to investigate the fate of the small
Dirac eigenvalues at $T=T_c$ analytically in random matrix models, as
was discussed at the end of Sec.~\ref{UniversalityProofs}.  There
are two recent lattice studies of this question \cite{Farc99b,Dam00}.
Some 
of the theoretical expectations were confirmed qualitatively, e.g.
the gap in the eigenvalue distribution above $T_c$ and the agreement
of the eigenvalue distribution in this region with the soft-edge
predictions of RMT.  However, further studies are needed to clarify
the situation at $T=T_c$.  We are looking forward to
upcoming work in this area.

\subsection{The Thouless Energy and Beyond}
\label{BeyondThouless}

\begin{figure}
  \begin{center}
    \epsfig{figure=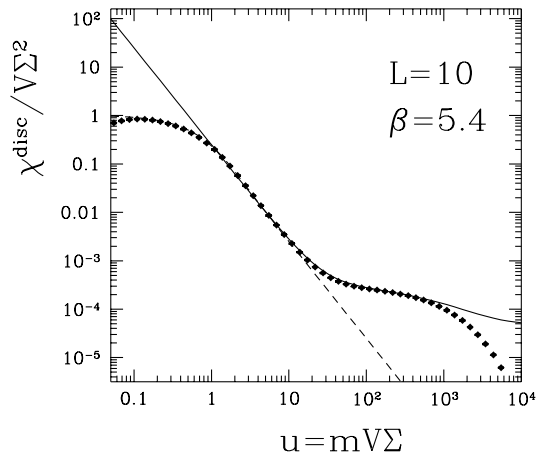,width=70mm}
    \caption{The data points represent the disconnected susceptibility
      computed on the lattice in quenched SU(3) with staggered
      fermions ($V=10^4$, $\beta=5.4$).  The solid line is the
      prediction of chPT, the dashed line the prediction of chRMT.
      (The dashed line is hidden by the data points for
      $u<10$.)  (From Ref.~\cite{Berb99}.)}
    \label{chPT}
  \end{center}
\end{figure}

As discussed above, the Dirac spectrum is described by chRMT
--- or, equivalently, by the zero-mode approximation of the low-energy
effective theory, --- only below the so-called Thouless energy $E_c$.  The
theoretical prediction for this quantity \cite{Verb96,Trento},
\begin{equation}
  E_c\sim\frac{F^2}{\Sigma\sqrt{V}}\:,
\end{equation}
has been confirmed quantitatively in the instanton liquid model
\cite{Osbo98a,Garc00} and on the lattice \cite{Berb98c,Gock98,Guhr99}.
To 
compute the Dirac spectrum beyond the Thouless energy, the calculation
must include the kinetic
terms in the low-energy effective theory, as discussed in
Sec.~\ref{sec:effective}.  The results of 
such an analysis can again be tested by lattice simulations.  This has
been done for staggered fermions \cite{Berb99}.  Note that in this
case, the effective low-energy theory must be modified to take into
account the
symmetries of staggered fermions at finite lattice spacing.
A convenient quantity to test the theory
is the disconnected scalar susceptibility, defined by
\begin{equation}
  \chi^{\mathrm {disc}}(m)
    =\frac{1}{N}\left\langle\sum_{k,l=1}^N \frac{1}
      {(i\lambda_k+m)(i\lambda_l+m)}\right\rangle
     -\frac{1}{N} \left\langle\sum_{k=1}^N\frac{1}
      {i\lambda_k+m}\right\rangle^2 \:.
\end{equation}
This quantity can also be computed in chRMT and in chiral perturbation
theory (chPT).  The result
of chRMT is expected to describe the data up to $E_c$.  The result of
chPT should describe the data beyond $E_c$ but is expected to break
down on the scale of the smallest eigenvalue, i.e.\ for $m\sim
1/V\Sigma$.  These expectations are confirmed by the lattice analysis
(see Fig.~\ref{chPT} for an example).

\section{SCHEMATIC MODELS}
\label{models}

We wish to emphasize again that
there are two different types of applications of RMT
to physical problems. RMT may be applied as an exact theory for
correlations of eigenvalues, as discussed in the first
part of this review, or it may serve as a schematic model
for chaos and disorder in physical systems. 
Below, we introduce a random matrix
model for the chiral phase transition at nonzero chemical potential
and temperature. Random matrix theories as schematic models are mostly
used for a description of nonuniversal phenomena.  This does not
exclude that they describe universal fluctuations of the eigenvalues
as well.  Typically, a random matrix model describes both universal
and nonuniversal properties.
Well-known examples are the Anderson model for localization phenomena
\cite{Ande58} and the use of random matrix theory in quantum gravity
\cite{Ginsparg,Dalmazi}.  

We argued in Sec.~\ref{sec1.1} that chiral symmetry breaking can
be understood in terms of the stiffness of the Dirac spectrum
resulting from interactions of the strong color force.  Because spectral
stiffness is a characteristic feature of RMT, it is
natural to describe the chiral phase transition in terms of a
random matrix model.  Several different types of schematic random
matrix models have been introduced. Here we discuss models for the
chiral phase transition at nonzero temperature $T$, for the chiral
phase transition at nonzero chemical potential $\mu$, and for the
phase diagram of QCD.  We also discuss random matrix models for
different types of fermions at $\mu>0$.

\subsection{Chiral Random Matrix Models for the Chiral Phase
  Transition at Nonzero Temperature}
\label{modelT}

The original motivation for introducing chiral random matrix models
was to obtain a better understanding of the QCD Dirac spectrum for
temperatures around the critical temperature for the chiral phase
transition \cite{JV,tilot}. The idea is to split the Dirac operator
into the time derivative and a remainder that will be replaced by a
chRMT,
\begin{equation} 
  {\cal D} = \gamma_0 \partial_0 + R\:.  
\end{equation} 
In a chiral basis with time dependence given by $\exp [i(2n+1)\pi
T\tau]$, where $\tau$ is the Euclidean time, the first term in ${\cal
  D}$ is diagonal and is given by a direct sum of Matsubara
frequencies $\omega_n = (2n+1)\pi T$. The simplest model for ${\cal
  D}$ is obtained by replacing the diagonal matrix with positive
Matsubara frequencies by the identity matrix times an effective
Matsubara frequency $t$, and replacing the diagonal matrix with negative
Matsubara frequencies by the opposite effective Matsubara frequency.
After a suitable basis change, this model can be written (in the sector
of zero topological charge) as \cite{JV,tilot}
\begin{equation} 
  Z_{N_f}(m) = \int DW {\det}^{N_f}({\cal D}+m) 
  e^{-\frac12 N\Sigma^2 \Tr (W^\dagger W)}\:,
  \label{zrandomt}
\end{equation}
where
\begin{equation}
  {\cal D} = \left (\begin{array}{cc} 0 & iW +it\\
      iW^\dagger +it& 0 \end{array} \right )\:.
  \label{diracop}
\end{equation}
We restrict ourselves to the unitary case ($\beta =2$) with a complex
matrix $W$ of dimension $N/2$.  The integral is over the real and
imaginary parts of the matrix elements of $W$.  For simplicity, we 
consider only a Gaussian probability distribution.  The normalization is
such that the parameter $\Sigma$ is equal to the magnitude of the
chiral condensate at zero temperature,
\begin{equation}
  \Sigma=\lim_{m\to0}\lim_{N\to\infty}\frac 1{NN_f} 
  \partial_m \left.\log Z_{N_f}(m) \right|_{t=0}\:.
\end{equation}

\begin{figure}
  \begin{center}
    \epsfig{file=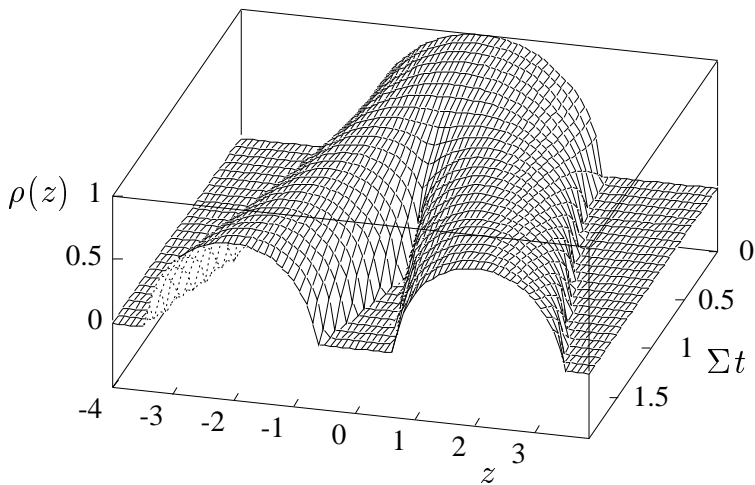,width=100mm} 
    \caption{The spectral density of the QCD Dirac operator as
      a function of an effective Matsubara frequency $t$ that models
      the effect of temperature.  (From Ref.~\cite{mishat}.)}
    \label{mishaspants}
  \end{center}
\end{figure}

It can be shown that the effect of the fermion determinant on the
macroscopic spectral density is subleading in $N_f/N$, so that
$\rho(\lambda)$ can be calculated in the quenched limit. By expanding the
resolvent of the Dirac operator, defined by
\begin{equation}
  G(z) = \frac 1N\:\Tr \left \langle  \frac 1{z-{\cal D}}
  \right\rangle\:,
\end{equation}
in a geometric series of the random matrix, it can be shown that
$g=G/\Sigma^2$ satisfies a cubic equation \cite{JV,senert,mishat},
\begin{equation}
  g^3 -2zg^2 +g(z^2-t^2 + 1/\Sigma^2) - z/\Sigma^2 = 0\:.
  \label{cubic}
\end{equation}
A variant of the method by which this equation was derived in
Refs.~\cite{JV,senert} is sometimes referred to as the Blue's function
method \cite{zeeblue}.  The spectral density, obtained from
the discontinuity of the resolvent, is shown in Fig.~\ref{mishaspants}.
We observe that chiral symmetry is broken up to $t = 1/\Sigma$ with a
chiral condensate given by
\begin{equation}
  \label{squareroot}
  G(T,z\rightarrow 0) = \Sigma\sqrt{1-(\Sigma t)^2}\:.
\end{equation}
Above this temperature, the spectrum splits into two disconnected
regions.
Since this model has no spacetime dependence, it should
not be a surprise that all critical exponents are given by their
mean field values.  {}From Eq.~(\ref{squareroot}), we see that $\beta
=1/2$.  For $t\to1/\Sigma$, we find
\begin{equation}
  G \propto z^{1/3}\:,
\end{equation}
resulting in another critical exponent $\delta = 3$. The spectral
density at the critical point follows by taking the discontinuity of
$G$ across the imaginary axis, see Eq.~(\ref{discon}), and is thus
given by
\begin{equation}
  \rho(\lambda) = \frac{N}{\pi} \lambda^{1/3}\:.
\end{equation}
The unfolded eigenvalues in this case are given by
\begin{equation}
  x_k = \frac {3N}{4\pi} \lambda_k^{4/3}\:,
\end{equation}
and a nontrivial scaling limit at $t=1/\Sigma$ is obtained by
introducing the microscopic variable \cite{Hikami-Brezin,Jani98a}
\begin{equation}
  z = \lambda N^{3/4}\:.
  \label{scalingt}
\end{equation}
Starting from the exact analytical result derived in Ref.~\cite{senert}, it
is possible to take the limit $N\rightarrow \infty$ with the scaling
relation (\ref{scalingt}) and to obtain an analytical result for the
microscopic spectral density at the critical point
\cite{Hikami-Brezin,Jani98a}.  We emphasize that this result is based
on mean field critical exponents and is thus not applicable to QCD
with nontrivial critical exponents (see also the discussions in
Secs.~\ref{UniversalityProofs} and \ref{sec:chphtr}).
 
As a last application of the partition function (\ref{zrandomt}), we
mention the explanation of the observation \cite{Chan95} that the
temperature at which chiral symmetry is restored is higher for gauge
field configurations 
with a nonzero $Z_3$-phase than for gauge field
configurations with zero $Z_3$-phase. The explanation in terms of the
random matrix model (\ref{zrandomt}) is that the lowest
Matsubara frequency for the nonzero $Z_3$-phases is shifted to a lower
value by the phase of the Polyakov loop, and thus the critical
temperature is higher \cite{mishat}.

\subsection{Chiral Random Matrix Models at Nonzero Chemical Potential}

\subsubsection{QCD PARTITION FUNCTION AT NONZERO CHEMICAL POTENTIAL}

The QCD partition function at nonzero temperature $T$ and chemical
potential $\mu$ is given by \cite{phase}
\begin{equation}
  Z(m,\mu,T) =\Tr e^{-\frac{H_{\rm QCD} -\mu N}T}
  =\sum_\alpha e^{-\frac{E_\alpha-\mu N_\alpha}{T}}\:,
\end{equation}
where $H_{\rm QCD}$ is the Hamiltonian of QCD with eigenvalues
$E_\alpha$ and $N$ is the quark number operator. At zero temperature,
only the states with $E_\alpha/N_{\alpha} < \mu$ contribute to the
partition function. We thus expect that the partition function is
independent of $\mu$ below a critical chemical potential $\mu_c$ given
by the lightest baryon mass per unit quark number.  Therefore, for
$\mu < \mu_c \approx m_N/3$ (where $m_N$ is the nucleon mass), the
baryon density remains zero and the chiral condensate is constant.

The quark chemical potential appears in the Lagrangian in the form
$\mu\psi^\dagger\psi=\bar\psi(\mu\gamma_0)\psi$ and is therefore
introduced in the Dirac operator by the substitution
\begin{equation}
  \partial_0 \rightarrow \partial_0 + \mu\:.
\end{equation}
This substitution destroys the Hermiticity properties of the QCD Dirac
operator, and the resulting complex phase of the fermion determinant
makes Monte Carlo simulations impossible.  Because of the success of
the quenched approximation in lattice QCD simulations at $\mu=0$, it
is tempting to ignore the fermion determinant in this case.
However, it was shown \cite{everybody,maria} that the critical
chemical potential for quenched simulations is determined by the pion
mass rather than the nucleon mass. Obviously, the basic physics of the
problem is not visible in the quenched approximation.  An analytical
understanding of this problem was first obtained by means of a random
matrix model for the Dirac operator of QCD at finite density
\cite{Misha}.

\subsubsection{A RANDOM MATRIX MODEL FOR QCD AT FINITE DENSITY}
\label{RMMmu}

A random matrix model for QCD at nonzero chemical potential model is
obtained by writing the Dirac operator as \cite{Misha}
\begin{equation}
  {\cal D}(\mu) = \mu \gamma_0 + {\cal R}
\end{equation}
and replacing ${\cal R}$ by a chiral random matrix ensemble. The partition
function is thus given by
\begin{equation}
  Z_{N_f}(m) = \int DW  {\det}^{N_f}({\cal D} +m)
  e^{-\frac12N\Sigma^2 \Tr(W^\dagger W)}
  \label{zrandommu}
\end{equation}
with a Dirac operator given by
\begin{equation}
  {\cal D} = \left (\begin{array}{cc} 0 & iW +\mu\\
      iW^\dagger +\mu& 0 \end{array} \right )\:.
  \label{diracopmu}
\end{equation}
For QCD with three or more colors (chUE) this Dirac
operator has no Hermiticity properties. Therefore, the
fermion determinant in Eq.~(\ref{zrandommu}) has a complex phase, and the
partition function cannot be simulated by Monte Carlo methods.

The partition function with Dirac operator (\ref{diracopmu}) can be
rewritten in terms of a $\sigma$-model. For $N_f= 1 $, the partition
function is particularly simple,
\begin{equation}
  Z(\mu) =\int d\sigma d\sigma^* [(\sigma+m)(\sigma^*+m) - \mu^2]^n
  e^{-n|\sigma|^2}\:, 
\label{zmusig}
\end{equation}
where we have set $\Sigma=1$ for ease of notation.  For $n=N/2\rightarrow
\infty$ the integrals can be evaluated by a saddle-point
approximation. In the chiral limit, we find that \cite{Misha}
\begin{equation}
  \bar \sigma = \left\{
    \begin{array}{ccccl}
      0 & {\rm for} & \mu >\mu_c & \Rightarrow & Z = \mu^{2n} \\
      \sqrt{1+\mu^2} & {\rm for} & \mu < \mu_c & \Rightarrow & 
      Z = e^{-n(\mu^2 + 1)}
    \end{array}\right.
\end{equation}
with $\mu_c$ given by the point where the two partition functions are
equal, i.e.\ $\mu_c^2 = \exp(-1-\mu_c^2)$, which is solved by $\mu_c
\approx 0.53$. The vacuum properties of this partition function depend
on $\mu$ for $\mu \le \mu_c$. The reason for this unphysical result
is discussed below in Sec.~\ref{failure}. As was demonstrated in
Ref.~\cite{Vanderheyden99} for $\beta =2$, this model does not show diquark
condensation.  However, a random matrix model that does show diquark
condensation was formulated in Ref.~\cite{Vanderheyden00}.

\begin{figure}
  \begin{center}
    \epsfig{figure=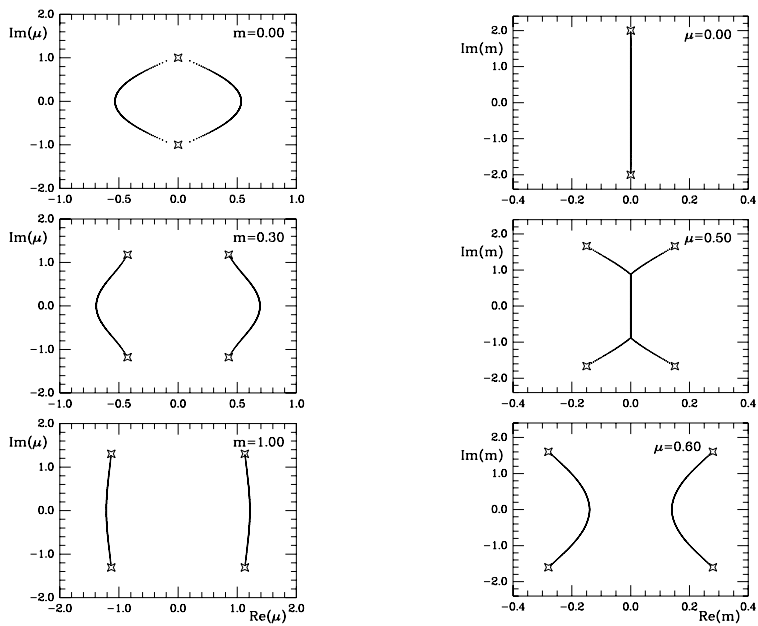,width=100mm}
    \caption{Zeros of the partition function in the complex $\mu$ and
      $m$ plane.  (From Ref.~\cite{Halaszyl}.)} 
    \label{ylfig}
  \end{center}
\end{figure}

The QCD partition function (\ref{zrandommu}) is a polynomial in $m$
and $\mu$ with coefficients that can be obtained analytically
\cite{Halaszyl}.  Its properties can be analyzed by means of its zeros
in the complex chemical potential plane and the complex mass plane.
Figure~\ref{ylfig} shows a first-order phase transition at $\mu
\approx 0.53$. In the chirally restored phase, the cut on the imaginary
$m$ axis is no longer present. The figure also shows the points (stars) where
two solutions of the saddle-point equations of the $\sigma$-model
coincide. Indeed, the cuts end on these branch points.

In lattice QCD at finite density, the zeros of the QCD function in the
complex $\mu$ plane can be studied by means of the Glasgow method
\cite{Barbourqed}.  This method has been analyzed in terms of the
above random matrix model \cite{glasgow} with the conclusion that it
requires an exponentially large number of gauge field configurations
to obtain statistically significant results.

\subsubsection{ZERO TEMPERATURE LIMIT} 

The random matrix model discussed in the previous section has
attracted a great deal of interest, and more sophisticated versions of
this model have been proposed
\cite{Feinnh,Halaszyl,Osbornmu,Nowak,glasgow}.  One problem of the
model (\ref{zrandommu}) is that the chiral condensate is
$\mu$-dependent below the critical value of the chemical potential.
The correct zero temperature limit, with a chiral condensate that
remains constant up to the critical value of the chemical potential,
is obtained if all Matsubara frequencies are taken into account.  We
illustrate this with a one-dimensional lattice QCD model.

Writing the Dirac operator as
\begin{equation}
  {\cal D}(\mu) = \gamma_0 (\partial_0+\mu) + R\:, 
  \label{Dmur}
\end{equation}
the $\mu$-dependence of the fermion determinant can be gauged away by
means of a time-dependent gauge transformation, i.e.
\begin{equation}
  \gamma_0 (\partial_0+\mu) + R = e^{-\mu\tau}
  \left [\gamma_0\partial_0  + R \right ] e^{\mu\tau}\:.
\end{equation}
The $\mu$-dependence is now in the boundary conditions, but for
$T\to0$ we expect that they are not important and that the partition
function becomes independent of $\mu$.  The lesson is that in order to
obtain a condensate that is $\mu$-independent, one must treat the time
derivative exactly, or, in other words, all Matsubara frequencies
have to be taken into account. This was first worked out in detail for
lattice QCD models at nonzero chemical potential \cite{matstone}. Let
us discuss the strong coupling limit of a one-dimensional SU($N_c$)
lattice QCD model with Kogut-Susskind fermions \cite{Bilic}.  In that
case, the matrix elements of the discretized Dirac operator are given
by
\begin{equation}
  {\cal D}_{kl} = \delta_{k,l-1} U_{kl} e^{\mu} -
  \delta_{k,l+1} U^\dagger_{kl} e^{-\mu} \:,
\end{equation}
where the indices are modulo $N$, the number of lattice sites.
Antiperiodic boundary conditions result in an extra minus sign for
the matrix elements ${\cal D}_{1N}$ and ${\cal D}_{N1}$. The matrices
$U_{kl}$ are independent SU($N_c$) matrices. By using gauge invariance,
one can reduce this partition function to a single SU($N_c$) integral,
which can be performed analytically, resulting in the partition
function
\begin{equation}
  Z_N = 2 \cosh(N_c N \mu) +
  \frac{\sinh [(N_c +1) N \sinh^{-1}m]}{\sinh[ N \sinh^{-1} m]}\:.
\end{equation}
The number of lattice sites $N$ can be interpreted as the total number
of Matsubara frequencies. A sharp transition is obtained in the limit
$N \rightarrow \infty$. For example, the chiral condensate is given by
\cite{Bilic}
\begin{equation}
  \lim_{N\to\infty}\frac 1N \partial_m \log Z_N = \left\{
    \begin{array}{cl}
      N_c/\sqrt{1+m^2} & \hbox{for $\sinh \mu <m$}\:,\\
      0 & \hbox{for $\sinh \mu  > m$}\:.
    \end{array}\right.
\end{equation}
Similar results have been derived for a large-$d$ strong coupling
expansion of lattice QCD at finite density \cite{strong} (with $d$ the
Euclidean dimensionality). The failure of the quenched approximation
has been reproduced in such lattice models.  The effect of including
all Matsubara frequencies has also been investigated in a random
matrix model obtained by replacing the remainder $R$ in Eq.~(\ref{Dmur})
by a chiral random matrix.  After ensemble averaging, one obtains a
similar partition function with similar conclusions \cite{papp}.

\subsubsection{QUENCHING AT NONZERO CHEMICAL POTENTIAL}
\label{failure}

The failure of the quenched approximation at nonzero chemical
potential was first understood analytically in terms of chRMT by
Stephanov \cite{Misha}.  He showed that the quenched limit is the
limit $N_f \to 0$ of a partition function with the absolute value of
the fermion determinant,
\begin{equation}
  |\det({\cal D}(\mu) +m)|^{N_f}\:,
  \label{absdet}
\end{equation}
rather than
\begin{equation}
  [\det({\cal D}(\mu) +m)]^{N_f}\:.
\end{equation}
The absolute value of the fermion determinant can be written as 
\begin{equation}
  \det({\cal D}+m) \det({\cal D}^\dagger+m^*)= 
  \det \left(\matrix{iW+\mu & m \cr m & \!\!\!iW^\dagger+\mu}\right)
  \det\left(\matrix{iW^\dagger-\mu& m^*\cr m^*&\!\!\!iW-\mu}\right)\:.
  \label{dets}
\end{equation}
Writing the fermion determinant as a Grassmann integral, we observe
that the quenched partition function can be interpreted as a partition
function of quarks and conjugate antiquarks. Therefore, in addition
to the usual Goldstone modes, we have Goldstone modes consisting of a
quark and a conjugate antiquark \cite{Goksch,Misha}.  Such modes, with
the same mass as the usual Goldstone bosons, have a nonzero baryon
number. The critical chemical potential given by the mass of the
lightest particle with nonzero baryon number is thus $m_\pi/2$.  The
failure of the quenched approximation thus has an important benefit:
It allows us to write down the exact low-energy effective partition
function, which is discussed in the next section.  The product of
the two determinants in Eq.~(\ref{dets}) can be written as the determinant
of a single Hermitian matrix.  This procedure is known as
Hermitization \cite{Girko,Fein,Efetov}.

\subsubsection{QUENCHED DIRAC SPECTRA} 

A chRMT for quenched QCD at nonzero chemical
potential is obtained by replacing the determinant in
Eq.~(\ref{zrandommu}) by its absolute value.  This model can be solved
analytically in the large-$N$ limit \cite{Misha}.  However, we will
follow a different approach by analyzing the corresponding effective
theory \cite{many1}.  The advantage of this approach is that the
effective theory is as valid for quenched QCD as it is for the
quenched chiral random matrix model.

For non-Hermitian matrices, the eigenvalues are scattered in the
complex plane.  Using the fact that
\begin{equation}
  \partial_{z^*} \frac 1z = \pi \delta^2(z) \:,
\end{equation}
where the complex delta function is defined as $\delta^2(z) =
\delta(\Re z)\delta(\Im z)$, we find that the two-dimensional spectral
density is (up to a normalization constant) given by
\begin{equation}
  \rho(\lambda) = \left .\frac 1\pi \partial_{z^*} G(z)\right
  |_{z=\lambda} \:. 
  \label{BCC}
\end{equation}
The resolvent is defined as
\begin{equation}
  G(z) = \frac 1{N_f V} \partial_{z} \log Z\:,
  \label{Gz}
\end{equation}
and the quenched result is obtained in the limit $N_f = 0$ of the
partition function
\begin{equation}
  Z=\left\langle {\det}^{N_f}({\cal D}(\mu)+z) {\det}^{N_f}
  ({\cal D}^\dagger(\mu) + z^*)\right\rangle \:.
  \label{zpart}
\end{equation}
The product in Eq.~(\ref{zpart}) can be written as
\begin{equation}
  {\det}^{N_f}({\cal D}(\mu)+z) {\det}^{N_f}( {\cal D}^\dagger(\mu) + z^*) = 
  \det \mat iW{\bf 1}_{2N_f} +B_R& \zeta \\ \zeta & iW^\dagger {\bf 1}_{2N_f}
    +B_L \emat
\end{equation}
with
\begin{equation}
  B\equiv B_R=B_L = \mat \mu{\bf 1}_{N_f} & 0 \\ 0 & -\mu{\bf 1}_{N_f} 
\emat \quad {\rm and} \quad
  \zeta = \mat z{\bf 1}_{N_f} & 0 \\ 0 & z^*{\bf 1}_{N_f} \emat\:,
\end{equation}
where we have displayed the degeneracy in flavor space by means of the
identity matrices ${\bf 1}_{2N_f}$ and ${\bf 1}_{N_f}$.  For $z=\mu
=0$, the quenched partition function is thus invariant under ${\rm
  SU}_R(2N_f)\times {\rm SU}_L(2N_f)$. This symmetry is broken
spontaneously to the diagonal subgroup, ${\rm SU}_V(2N_f)$.  At low
energies, the effective partition function is therefore given by a
partition function of Goldstone modes parameterized by matrices
$U\in{\rm SU}(2N_f)$. 
The static effective Lagrangian is obtained from the requirement that
it should have the same symmetries as the underlying microscopic
partition function. The mass term was discussed in
Sec.~\ref{Finite-Volume}.  The chemical potential term remains
invariant under ${\rm SU}(2N_f)\times{\rm SU}(2N_f)$ transformations
if at the same time $B_R$ and $B_L$, now considered as independent
matrices, are transformed as
\begin{equation}
  B_R \rightarrow  U_R B_R U_R^{-1} \:, \qquad B_L \rightarrow U_L B_L
  U_L^{-1}\:. 
\end{equation}
The matrices $U$ in the Goldstone manifold transform as $U \to U_R U
U_L^{-1}$. 
To lowest nontrivial order in $\mu$, we can write two invariant
combinations,
\begin{equation}
  \Tr U B_L U^{-1} B_R \qquad {\rm and} \qquad \Tr BB\:.
\end{equation}
The coefficients follow from the conditions that the critical chemical
potential is equal to one third of the mass of the lightest baryon and
that the baryon density should vanish below $\mu_c$.  For the static
limit of the effective partition function, we find
\begin{equation}
  Z = \int_{U\in {\rm SU}(2N_f)} DU e^{-\frac {F^2 V}4\Tr[ U, B]
    [ U^{-1}, B] + \frac12\Sigma V \Tr (MU + M^\dagger U^{-1})}\:,
\label{zeffspec}
\end{equation}
with the mass matrix given by
\begin{equation}
  M = \mat z {\bf 1}_{N_f} & 0 \\ 0 & z^* {\bf 1}_{N_f} \emat\:.
\end{equation}
This partition function can be derived more elegantly by means of 
a local gauge invariance principle \cite{KST,KSTVZ}.
For real masses $z$ the Dirac operator
satisfies the relation 
${\rm det}({\cal D}^\dagger(\mu) + z^*)=
{\rm det}({\cal D}(-\mu) + z)$, so that the partition function 
defined in Eq.~(\ref{zpart}) and the corresponding low-energy effective
partition function given in Eq.~(\ref{zeffspec}) can  be
interpreted as the partition function of QCD at finite isospin density
\cite{misha-son}.

Below the critical chemical potential, $\mu < m_\pi/2$, only the
vacuum state contributes to the partition function, so that
\begin{equation}
  Z = e^{\Sigma V N_f(z+z^*)}\:.
\end{equation}
In terms of the effective theory, the saddle point is at $U =\bf 1$.  The
resolvent (\ref{Gz}) is given by
\begin{equation}
  \label{Gzconstant}
  G(z) = \frac 1{N_f V} \partial_{z} \log Z = \Sigma\:.
\end{equation}
Since $G(z)$ does not depend on $z^*$, it follows from Eq.~(\ref{BCC})
that the spectral density is zero in the region where
(\ref{Gzconstant}) holds.  Because $m_\pi^2=(z+z^*)\Sigma/F^2$, see
Eqs.~(\ref{Gmass1}) through (\ref{Gmass3}), the condition $\mu <
m_\pi/2$ means that the spectral density vanishes everywhere except in
a strip
\begin{equation}
  |\Re z| < \frac{2\mu^2 F^2}\Sigma\:.
\end{equation}
For $z$ inside this strip, the Goldstone modes contribute to the
partition function.  In terms of the effective partition function, the
saddle point rotates away from $U=\bf 1$, leading to a nonvanishing diquark
condensate.  The rotation of the saddle point is a generic feature of
low-energy effective partition functions of non-Hermitian field theories
\cite{supernonh,Efetovnh,KST,KSTVZ,efetovrot}.

These results are in qualitative agreement with spectra of the
quenched lattice QCD Dirac operator at $\mu \ne 0$ \cite{everybody}.
The correlations of the complex eigenvalues are not so well
understood, but the first lattice QCD calculations \cite{Tilomu} seem
to confirm the theoretical expectations
\cite{Ginibre,fyodorov,fyodorovpoly,Forrestc}.

\subsection{Phase Diagram for the QCD Partition Function}

The random matrix models at nonzero temperature and nonzero chemical
potential introduced in the previous sections can be merged into a
single schematic chRMT model for the chiral phase transition
\cite{phase},
\begin{equation}
  Z = \int DW {\det}^{N_f} \left ( \begin{array}{cc} m & iW + iC\\
      iW^\dagger +iC& im \end{array}\right )
  e^{-\frac12N\Sigma^2\Tr WW^\dagger}\:,
  \label{Zmut}
\end{equation}
where $C$ is a diagonal matrix with $C_{kk}= t -i\mu$ for one half of the
diagonal elements and $C_{kk}=-t-i\mu$ for the other half.  In the
following, we set $\Sigma=1$ for simplicity.

This random matrix partition function can be rewritten as a
$\sigma$-model.  For the case $N_f = 1$ it is given by
\begin{equation}
  Z(t,\mu) = \int d\sigma e^{-\frac{N}{2}\Tr\Omega (\sigma)}\:,
\end{equation}
where
\begin{eqnarray}
  \Omega(\sigma) &\!\!\!=\!\!\!& \sigma \sigma^* 
  -\frac12\log [(\sigma+m)(\sigma^* + m) - (\mu+it)^2]\nonumber \\
  &&\phantom{\sigma \sigma^*}-\frac12\log [(\sigma+m)(\sigma^* + m) 
  - (\mu-it)^2]\:.
\end{eqnarray}
For $t = 0$ this is exactly the $\sigma$-model discussed in
Sec.~\ref{RMMmu}.  At the saddle point, the chiral condensate is given
by the expectation value of $\sigma$. For $m= 0$ the saddle-point
equation is of fifth order in $\sigma$,
\begin{equation}
  \sigma[\sigma^4 - 2(\mu^2 -t^2 +\frac 12)\sigma^2 
  + (\mu^2+t^2)^2 +\mu^2 -t^2] =0\:.
  \label{saddlemut}
\end{equation}
The critical points occur where one of the solutions of the quartic
equation merges with the solution $\sigma = 0$, i.e.\ along the curve
$(\mu^2+t^2)^2 +\mu^2 -t^2 =0$. At the tricritical point, three
solutions merge. This happens if in addition $\mu^2 -t^2 +\frac 12=0$.

\begin{figure}
  \begin{center}
    \epsfig{file=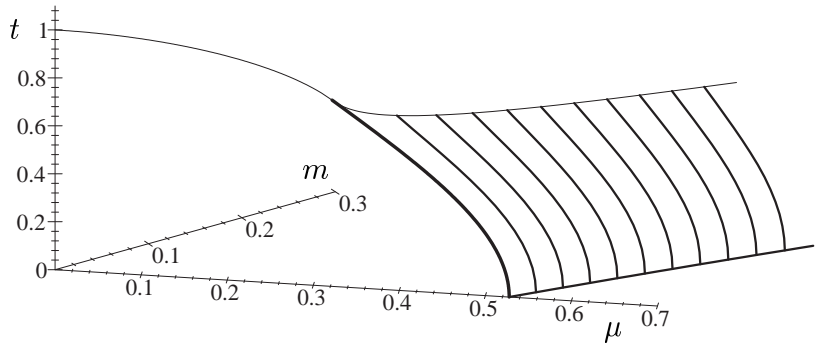,width=120mm}
    \caption{Phase diagram of QCD with two light flavors of mass $m$
      as calculated from the random matrix model.  The almost parallel
      curves on the wing surface are cross sections of this surface
      with $m=$constant planes.  (From Ref.~\cite{phase}.)}
      \label{fig:3dRM}
  \end{center}
\end{figure}  
Figure~\ref{fig:3dRM} shows the phase diagram in the $\mu t
m$-space. In the $m=0$ plane, we observe a line of second-order phase
transitions and a line of first-order phase transitions. They join at
the tricritical point.  Also joining at the tricritical point is a
line of second-order phase transitions in the $m$-direction, which is
the boundary of the plane of first-order transitions in $\mu
tm$-space. This line is the collection of end points of lines of
first-order phase transitions.  We expect that the critical exponents
for such a liquid-gas transition are given by the three-dimensional Ising
model.  The tricritical point was also found in a Nambu--Jona-Lasinio
model \cite{italians,krish3}.

Because the saddle-point equation (\ref{saddlemut}) is of fifth order
in $\sigma$, the critical properties of the random matrix model
(\ref{Zmut}) are very similar to those of a $\phi^6$-theory.  The
critical dimension at the tricritical point of such theories is three, so
that mean field theory, and therefore RMT, describes
the correct critical behavior at this point.  This random matrix model
can be considered as the matrix equivalent of a Landau-Ginzburg
functional.  The advantage over using the standard Landau-Ginzburg
theory is that in this case the spectrum of the Dirac operator is also
accessible. This allows us to study the critical properties of the
Dirac eigenvalues. For example, for $\mu = 0$ we have found that the
distribution of the smallest nonzero eigenvalue of the Dirac operator
may serve as an order parameter for the chiral transition \cite{JV}.

\subsection{Random Matrix Triality at $\mu \ne 0$}

In previous sections, we have shown that the pattern of chiral symmetry
breaking and the correlations of the Dirac eigenvalues are related to
the anti-unitary symmetries of the Dirac operator. Since the chemical
potential occurs only in the combination $\partial_0 + \mu$, the
anti-unitary symmetries at $\mu \ne 0$ are the same as for zero
chemical potential. Thus for QCD with two colors and fermions in the
fundamental representation the Dirac operator is
real ($\beta =1$), whereas for QCD with adjoint fermions the Dirac
operator is quaternion real ($\beta =4$). In the first case, the Dirac
operator has the structure \cite{Osbornmu}
\begin{equation}
  {\cal D} = \mat 0 & W+ \mu \\ -W^T +\mu & 0 \emat \quad \hbox{with
  $W$ real}.
\end{equation}
In the second case, ${\cal D}$ is given by
\begin{equation}
  {\cal D} = \mat 0 & W+ \mu \\ -W^\dagger +\mu & 0 \emat \quad 
  \hbox{with $W$ quaternion real}.
\end{equation}
In this case, the quaternion real matrix elements of $W$ satisfy the
reality relation $W_{kl}^* =\sigma_2 W_{kl} \sigma_2$. In both cases,
it is easy to show that the fermion determinant is real.
Furthermore, because of
\begin{eqnarray}
  {\det}^{N_f} W\,{\det}^{N_f} W^T = {\det}^{2N_f} W &&
  \hbox{for $\beta = 1$},\\
  {\det}^{N_f} W\,{\det}^{N_f} W^\dagger = {\det}^{2N_f} W &&
  \hbox{for $\beta = 4$},
\end{eqnarray}
the flavor symmetry group is enhanced to U($2N_f$) in both cases (for
$\nu=0$).\footnote{For $\beta=4$ and $\mu=0$, the Dirac operator
  satisfies the relation $(C{\cal D})^T=-C{\cal D}$ whereas
  $(C\mu\gamma_0)^T=C\mu\gamma_0$ (with $C$ the charge conjugation
  matrix).  Therefore, for $\beta =4$ and $\mu \ne 0$, the Pfaffian of
  the Dirac operator is not defined and the fermion determinant
  ${\det}^{N_f/2}({\cal D}+\mu\gamma_0)$ cannot be expressed as an
  integral over $N_f$ Majorana fermions with flavor symmetry U($N_f$),
  as discussed in Sec.~\ref{anti-unitary}.  Instead, we write
  ${\det}^{N_f}({\cal D}+\mu\gamma_0)$ as an integral over $N_f$ Dirac
  fermions with flavor symmetry U($2N_f$).}
The full flavor symmetry is broken spontaneously to
Sp($2N_f$) and O($2N_f$), respectively, and is broken in the same way
by the mass term. The chemical potential breaks the symmetry according
to ${\rm U}(2N_f) \rightarrow {\rm U}(N_f)\times {\rm U}(N_f)$ in both
cases.  For both 
$\beta =1 $ and $\beta=4$ the pseudoreality of the Dirac operator
leads to Goldstone modes with a nonzero baryon number in the same way
we have seen for the phase quenched partition function. An exact
low-energy effective partition function valid to lowest order in $m$
and $\mu^2$ can be written down \cite{KSTVZ}.  For an elaborate
discussion of the symmetries, their breaking pattern, and the
low-energy effective theory in both cases, see Ref.~\cite{KSTVZ}.

For an even number of flavors, dynamical Monte Carlo simulations are
possible for $\beta =1$ and $\beta =4$, though not for
$\beta=2$. However, quenched simulations have been 
performed for all three classes.  A cut along the imaginary axis below
a cloud of eigenvalues was found in instanton liquid simulations
{\cite{Thomas}} for $N_c =2$ at $\mu \ne 0$, which corresponds to
$\beta =1$. In lattice QCD simulations with staggered fermions for
$N_c = 2$ {\cite{baillie}}, a depletion of eigenvalues along the
imaginary axis was observed, whereas for $N_c=3$ the eigenvalue
distribution did not show any pronounced features \cite{everybody}.
\begin{figure}
  \begin{center}
    \epsfig{file=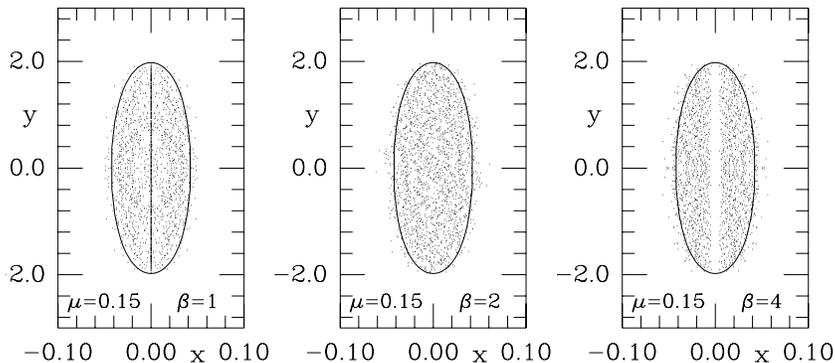,width=110mm,angle=0}
    \caption{Scatter plot of the real ($x$) and imaginary ($y$) parts 
      of the eigenvalues of the random matrix Dirac operator at
      nonzero chemical potential.  The values of $\beta$ and $\mu$ are
      given in the figures.  Full curves show the analytical
      result for the boundary.  The line along the imaginary axis for
      $\beta=1$ represents the accumulation of eigenvalues on this
      axis.  (From Ref.~\cite{Osbornmu}.)}
    \label{fig:scatter}
  \end{center}
\end{figure}

A chiral random matrix model for the Dirac operator at $\mu \ne 0$ in
each symmetry class is obtained by drawing the matrix elements of $W$
from a Gaussian probability distribution.  In the quenched
approximation, the spectral properties of the random matrix Dirac
operator of Eq.~(\ref{diracopmu}) can easily be studied numerically by
diagonalizing a set of matrices with the probability distribution
of Eq.~(\ref{zrandommu}).  Figure~\ref{fig:scatter} shows results
{\cite{Osbornmu}} for the eigenvalues of a few $100\times 100$
matrices for $\mu = 0.15$ (dots). The solid curves represent the
analytical result for the boundary of the domain of eigenvalues
derived in {Ref.~\cite{Misha}} for $\beta =2$. However, the analysis
can be extended {\cite{Osbornmu}} to $\beta = 1$ and $\beta
=4$, and with the proper scale factors, the solution is identical.
                                        
For $\beta =1$ and $\beta = 4$ we observe exactly the same structure
as in the previously mentioned (quenched) QCD simulations.  There is an
accumulation of eigenvalues on the imaginary axis for $\beta = 1$ and
a depletion of eigenvalues along this axis for $\beta = 4$.  The
depletion can be understood as follows. For $\mu = 0$ all eigenvalues
are doubly degenerate. This degeneracy is broken at $\mu\ne 0$, which
produces the observed repulsion between the eigenvalues.

The number of purely imaginary eigenvalues for $\beta = 1$ scales as
$\sqrt N$ and is thus not visible in a leading-order saddle-point
analysis.  Such a $\sqrt N$ scaling is typical for the regime of weak
non-Hermiticity first identified by Fyodorov et al.
{\cite{fyodorov}}.  Using the supersymmetric method of RMT,
Efetov {\cite{Efetovnh}} obtained the $\sqrt N$ dependence
analytically.  The case $\beta=4$ was also analyzed analytically
\cite{efetovsym}, with results that are in complete agreement with
the numerical simulations.  Obviously, more work is needed in
order to arrive at a complete characterization of universal features
{\cite{fyodorovpoly}} of the spectrum of non-Hermitian operators.

\section{RELATED MODELS AND RELATIONS TO OTHER \mbox{FIELDS}}
\label{otherfields}

Random matrix theory has been used extensively in many different
fields including nuclear physics, atomic and molecular physics,
condensed matter physics, quantum chaos, quantum gravity, and
mathematical physics.  Before reviewing a few applications that might
have some relation to QCD, we briefly discuss another class of matrix
models for the QCD partition function.  For a more comprehensive
review of the material in this section, see
Refs.~\cite{Guhr98,Beenreview,Montambaux,eeee,mironov}.

\subsection{One-Plaquette Models of Lattice QCD}

One class of matrix models for QCD are lattice QCD partition functions
on a $2^d$ lattice. The simplest model in this class is the
one-plaquette partition function for pure gauge theory in $d=2$. This 
model is known as the Br\'ezin-Gross-Witten model \cite{GW} and is
defined by
\begin{equation}
  Z(J,J^\dagger) = \int_{U \in {\rm U}(N)} DU e^{\Tr (JU^\dagger
  +J^\dagger U)}\:, 
\end{equation}
where the integral is over the Haar measure of U($N$) and $J$ is an
arbitrary complex source term. We have shown that this
partition function can be expressed as a determinant of modified
Bessel functions \cite{brower,Leut92,seneru}. For $J$ a multiple of
the identity (with the interpretation of $J\sim1/g^2$, where $g$ is
the Yang-Mills coupling constant), this model was solved analytically
in the large-$N$ limit by Gross and Witten.  They found a third-order
phase transition at a critical value of $J$. A large-$N$ chiral phase
transition is obtained by extending this model with a fermion
determinant \cite{Chanran}.  Despite considerable effort, the
generalization of this model to more than two dimensions, which is
known as the Eguchi-Kawai model \cite{EK}, could not be solved
analytically \cite{Das}.

The Br\'ezin-Gross-Witten model \cite{GW} can be rewritten as a
generalized Kontsevich model, which has received a great deal of
attention in the theory of exactly integrable systems
\cite{dijkgraaf,mironov}.  There are many other relations between RMT
and the theory of exactly solvable systems. Among others, the
asymptotic properties of correlation functions can be derived by means
of conformal field theory \cite{Kostov}.  An interesting overview in
the context of string theory is given in Ref.~\cite{dijkgraaf}.
 
Another class of related models is the Kazakov-Migdal model
\cite{Kazakov}, which is also known as induced QCD.  In its simplest
form it is defined by
\begin{equation}
  Z = \int DH DU e^{ \sum_x  \Tr H^2(x)+ \sum_{x,\mu}
    \Tr[U(x) H(x) U^\dagger(x) H(x+\mu)]} \:,
\end{equation}
where the $H(x)$ are Hermitian matrices and the $U(x)$ are unitary
matrices.  The sum is over a lattice in $d$ dimensions.  A
zero-dimensional form of this model was proposed as a model for the
transition between Poisson statistics and Wigner-Dyson statistics
\cite{Moshe,Garc00}.

\subsection{Universal Conductance Fluctuations in Disordered
  Mesoscopic Systems} 
\label{sec:ucf}

In condensed matter physics, a mesoscopic system is a system whose
linear size is larger than the elastic mean free path of the electrons
but smaller than the phase coherence length, which is essentially the
inelastic mean free path.  A typical size is about 1~$\mu$m.  The
conductance $g$ of mesoscopic samples is closely related to their
spectral properties.  Using a scaling block picture, Thouless found
that in the diffusive regime, the conductance is given by
$g=E_C/\Delta$, where $E_C/\hbar$ is the inverse diffusion time of an
electron through the sample and $\Delta$ is the mean level spacing
\cite{Thou74}.  This can be rewritten as $g=\langle N(E_C)\rangle$,
where $\langle N(E)\rangle$ is the mean level number in an energy
interval $E$.  Thus the variance, $\langle\delta g^2\rangle$, of the
conductance is related to the number variance, $\Sigma^2$, of the
energy levels \cite{altshuler}.

Low-temperature experiments have been performed in which the
conductance of mesoscopic wires was measured as a function of an
external magnetic field.  The observed fluctuations in $g$ are of the
order of $e^2/h$, independent of the details of the system (shape,
material, etc).  These are the so-called universal conductance
fluctuations \cite{Wash86}.  One can understand this phenomenon
qualitatively by estimating the number fluctuations of the electron
levels using RMT results.  Both the magnitude of the fluctuations and
their universality can be obtained through the transfer matrix
method.  An interesting numerical result is that the density of
eigenvalues of the transmission matrix in the Hofstadter model for
universal conductance fluctuations can be described in terms of the
microscopic spectral density of the chUE \cite{Hofstad,Slevin}.

\subsection{Anderson Localization}
\label{sec:anderson}

In more than two dimensions, a good conductor becomes an insulator
when the disorder becomes sufficiently strong. This phenomenon is
called Anderson localization.  In the localized phase, the wave
function of the electron is not described by Bloch waves, but by a
localized form that decays exponentially,
\begin{equation}
  \psi(r) \sim e^{-  r/{L_c}}\:.
\end{equation}
The length scale $L_c$ is known as the localization length.  This
phenomenon was first described by the Anderson model \cite{Ande58},
which is a hopping model with a random potential on each lattice
point. The dimensionality of the lattice plays an important role.  It
has been shown that in one dimension all states are localized. The
critical dimension is two, whereas in three dimensions there is a
delocalization 
transition at an energy $E_L$. The states below $E_L$ are localized
whereas the states above $E_L$ are extended, i.e.\ with a wave
function that scales with the size of the system.  The eigenvalues of
the localized states are not correlated, and their correlations are
described by the Poisson distribution.

An interesting question is whether Dirac eigenfunctions can be
exponentially localized. Parisi argued that localized
states can only occur in quenched systems \cite{Parisi}.  In QCD this
argument goes as follows.  If the eigenfunctions were spatially
localized, the eigenvalues would be uncorrelated, and there would be
no repulsion between the eigenvalues.  Because of the fermion
determinant and the measure in Eq.~(\ref{zeig}), the eigenvalues would
be repelled from the origin.  Since there would be no mechanism to
compensate for this repulsion, $\rho(0)/V$ and the chiral condensate
would be zero. Therefore, if chiral symmetry is spontaneously broken,
the eigenfunctions of the Dirac operator must be spatially extended.
Indeed, this has been found for the wave functions of the Dirac
operator for a gauge field ensemble given by a liquid of instantons
\cite{Osbo98a} as well as in lattice QCD \cite{Berb00}.

In the extended domain, the situation is more complicated. An important
energy scale is the Thouless energy \cite{Thou74}, which is related to
the diffusion time of an electron through the sample (see
Sec.~\ref{sec:ucf}). With that diffusion time given by $L^2/D$ (the
diffusion constant is denoted by $D$), the Thouless energy is
\cite{Altshuler}
\begin{equation}
  E_c = \frac{\hbar D}{L^2}\:.
\end{equation}
On time scales larger than $\hbar/E_c$, an initially localized wave
packet diffuses all over phase space.  If this wave function $\psi(t)$
at $t>\hbar/E_c$ is expressed as a superposition of eigenfunctions
$\phi_i(t=0)$, many of the overlaps $\langle\psi(t)|\phi_i\rangle$ are
nonzero.  Therefore we expect that for energy differences below $E_c$
the eigenvalues are correlated according to RMT.

The diffusive behavior of electrons is described by Goldstone modes,
called diffusons.  Classically, they satisfy a diffusion equation
\cite{Altshuler,Stone,Efetov}. The diffusion modes can be described in
terms of a $\sigma$-model, which has the same structure as the
low-energy effective theory for the QCD partition function. The
Thouless energy corresponds to the energy scale below which the
fluctuations of the zero-momentum modes dominate the fluctuations of
the nonzero-momentum modes, i.e.\ the scale discussed in
Sec.~\ref{sec:domains}, below which the QCD Dirac spectra are given by
chRMT. The analysis of classically chaotic quantum systems 
has made it clear that spectra are described by RMT if and only if the
corresponding classical motion is chaotic \cite{Bohi84,Seli84}.  If we
interpret the Dirac operator as a Hamiltonian in four Euclidean
dimensions plus one artificial time dimension, we thus
conclude that the classical motion of the quarks is
chaotic\footnote{There is an elaborate literature on the classical
  dynamics of gauge fields. For a discussion of this topic, see
  Ref. \cite{Biro}.}  \cite{Vminn} and therefore diffusive in four
Euclidean dimensions and one artificial time dimension, in agreement
with the interpretation of Goldstone modes as diffusion modes
\cite{Stone,Polonyi}. The equivalent of the diffusion constant can be
read off immediately by comparing the two $\sigma$-models, or
equivalently by comparing the scale of Eq.~(\ref{spectralrange}) with the
Thouless energy, and is given by $D \sim F^2/\Sigma$
\cite{Osbo98a,Jani98}.  Quantum mechanically, the Thouless energy can
be interpreted as the spreading width of the exact eigenstates over
the states of the noninteracting Hamiltonian \cite{spreading,Guhra}.

\subsection{Non-Hermitian Models with Disorder}

In the past few years, several non-Hermitian models with disorder have
been introduced in the literature. We mention only the simplest model,
which was motivated by the study of flux-line pinning in
superconductors \cite{Miller,Hata96}. Its Hamiltonian is given by
\begin{equation}
  \label{HN}
  H=\frac{1}{2m}({\bf p}+i{\bf h})^2+V({\bf r})\:,
\end{equation}
where $V({\bf r})$ is a random disorder potential, {\bf p} is the
momentum operator, and $i{\bf h}$ is a constant imaginary vector
potential.  The new feature of such models is that a
localization-delocalization transition can occur even in one and two
dimensions.  Although the wave functions corresponding to real
eigenvalues remain localized, the wave functions of complex
eigenvalues can be extended. For a detailed discussion of such models,
see Refs.~\cite{Fein,piet}.

The dependence on ${\bf h}$ in the Hamiltonian $H$ occurs only in the
combination ${\bf p} + i{\bf h}$, i.e.\ in exactly the same way as the
chemical potential in the QCD Dirac operator.  There are two important
differences from QCD: $(a)$ the operator $H$ does not have a
chiral structure, and $(b)$ the disorder is uncorrelated Gaussian and
quenched.  The connection between delocalization in the Hatano-Nelson
model (\ref{HN}) and diquark condensation in quenched QCD at nonzero
chemical potential is not yet understood and deserves further
attention.

After the initial work of Ginibre \cite{Ginibre}, non-Hermitian RMT
has received a great deal of attention in the mathematics literature
as well. Probably the best overview of results in this area is in the
book by Girko \cite{Girko}. 

\subsection{Andreev Scattering}

The term Andreev scattering (or Andreev reflection) refers to a
process in which an electron hits the interface between a normal metal
and a superconductor and is reflected as a hole with the opposite
momentum (or vice versa) \cite{Andr64}.  Stated differently, two
electrons can tunnel through the interface so that a Cooper pair is
added to the superconducting condensate (or removed from it).  This
process can be described in a microscopic mean field model by the
Bogoliubov-deGennes Hamiltonian, which can be written in matrix form as
\begin{equation}
  H=\left(\matrix{A&B\cr B^\dagger&-A^T}\right)\:,
\end{equation}
where $A$ ($-A^T$) represents the Hamiltonian for particles (holes)
and $B$ represents the pairing field.  The requirement that $H$ be
Hermitian means that $A$ must be Hermitian.  
If the system under consideration is invariant under time reversal or
spin rotation or both, there may be additional symmetries.  If the
system is invariant under spin rotation, $B$ is symmetric (in this
case we consider the Hamiltonian in the space of spin-up states only).
If, in addition, the system is invariant under time reversal, the
elements of $H$ are real.  Otherwise, they are complex.  On the other
hand, if the system is not invariant under spin rotation, $B$ is
antisymmetric.  In this case, if the system is invariant under time
reversal, the elements of $H$ are quaternion real.  Otherwise, they
are complex.  This classification establishes four new random matrix
ensembles \cite{Oppermann,Altland}.  The microscopic spectral
correlations of 
these ensembles are identical to those of the chiral ensembles if the
parameters (the Dyson index $\beta$ and the number of massless flavors
$N_f$) are chosen appropriately.

\subsection{Mathematical Physics and Quantum Gravity}

RMT has received a great deal of attention as a
problem in mathematical physics. We mention the relation between
universal behavior in RMT and the asymptotic behavior of orthogonal
polynomials \cite{Fox64,nagao,Brez93,Eynard,Deo}, the relation between
the classification of random matrix ensembles and the Cartan
classification of symmetric spaces \cite{class}, and the theory of
Riemannian superspaces \cite{class}.  Various methods for the
solution of random matrix problems have been proposed. We mention the
mapping to a gas of noninteracting fermions, the Coulomb gas method
\cite{Dyso62}, the Brownian motion method \cite{Dyso62}, the replica
method \cite{edwards}, the orthogonal polynomial method \cite{Mehta},
the supersymmetric method \cite{Efetov}, and the operator method
\cite{Ginsparg}.

Generally, the problem for $\beta=2$ is much simpler than for $\beta
=1$ and $\beta =4$. Nevertheless, a great deal of progress has been
made for $\beta =1$ and $\beta =4$. We mention relations between the
kernels of the correlation functions
\cite{Verb94,Naga95,Sene98,Wido98} and the relation between massless
and massive correlators \cite{akemann-kanzieper}.  Novel mathematical
methods have been developed for $\beta =1$ and $\beta =4$.  We 
mention only the skew-orthogonal polynomial method
\cite{Dyson-skew,Mehtaskew,nagao,Verb94} and the extension of an
operator method for $\beta =2$ \cite{brezinneuberger,tracysk,Sene98}.

There exists an elaborate literature on the random matrix formulation
of 2-$d$ quantum gravity (see e.g.\ the recent reviews by Abdalla et
al.~\cite{Dalmazi} and Di Franceso et al.~\cite{Ginsparg}). The idea
is that the partition function, which is a 
sum over all metrics, can be rewritten as a sum over random surfaces.
The sum over the dual graphs of a discretization of these random
surfaces can be rewritten in terms of a random matrix partition
function that can be analyzed with standard random matrix methods.

\section{CONCLUSIONS}

Chiral random matrix theory is the archetypal model for the
spontaneous breaking of chiral symmetry.  In chRMT, chiral symmetry is
spontaneously broken for any finite number of massless flavors, whereas
in QCD, breaking is believed to occur only below a certain
number of massless flavors, perhaps as few as four.  In this
review, we have studied chiral symmetry breaking from the perspective
of the eigenvalues of the QCD Dirac operator, where broken chiral
symmetry implies that the smallest eigenvalues are spaced as $1/V$.
The robustness of chiral symmetry breaking in chRMT can then be
understood as the crystallization of the eigenvalues by the long-range
random interactions.  The number of eigenvalues in a sequence
containing $N$ eigenvalues fluctuates on average only by $\sqrt{\log
  N}/\pi$ as opposed to $\sqrt N$ for uncorrelated eigenvalues. For
uncorrelated eigenvalues, we do not expect chiral symmetry breaking for
any number 
of massless flavors.  The conclusion is that chiral symmetry breaking
in QCD requires strongly correlated Dirac eigenvalues but not quite
as strongly correlated as in chRMT. To make this conclusion more
quantitative, we 
have formulated an effective theory for the QCD Dirac spectrum, which,
in addition to the usual Goldstone modes, contains Goldstone bosons of
spectral quarks that are ``ghost'' quarks introduced to probe the
Dirac spectrum.  Like the usual chiral Lagrangian, this
Lagrangian consists of a kinetic term and a mass term. An important
scale is the spectral mass for which the long-wavelength fluctuations
of these two terms are of equal order of magnitude. In the theory of
disordered condensed matter systems, this scale is known as the
Thouless energy.  For spectral masses below the Thouless energy, the
mass dependence is given by the zero-momentum sector of the effective
theory, and it coincides with that of chRMT in the limit of large
matrices.  The deviations from chRMT, given by the contributions of
the nonzero-momentum modes, increase the fluctuations of the
eigenvalues. For an increasing number of flavors, the position of the
smallest eigenvalues moves to the right, whereas the slope of the spectral
density increases with $N_f$.  However, chiral symmetry remains
broken for any value of $N_f$. The chiral condensate is simply a
parameter of the effective partition function.

The statistical properties of the QCD Dirac eigenvalues have been
investigated by numerous lattice QCD simulations, and the chRMT
predictions have been verified in great detail. The behavior of Dirac
spectra in the domain beyond the Thouless energy has been less well
studied, but in all cases, the predictions of the
effective theory for the Dirac spectrum agree well with lattice
simulations. 

Inspired by the successes of chRMT as an exact description of the
statistical properties of the QCD Dirac spectrum, we have constructed a
schematic model for the chiral phase transition.  Although this
approach does not provide rigorous results, it has been 
useful in advancing a qualitative understanding of the chiral phase
transition at both zero and nonzero chemical potential.  Examples are
the properties of the Dirac spectrum in the approach to the critical
temperature, the failure of the quenched approximation at $\mu>0$, and
the phase diagram of the QCD partition function in the $\mu Tm$ plane.
 
Finally, RMT encapsulates a duality between order and
chaos. Complete mixing of all degrees of freedom can be described by a
single effective degree of freedom. This is the progress we have
made toward an understanding of the complexity of the QCD vacuum.

\section*{Acknowledgments}

This work was supported in part by the US Department of Energy under
contracts DE-FG-88ER40388, DE-FG02-91ER40608, and DE-AC02-98CH10886.
We would like to acknowledge discussions and collaborations on the
subject of this review with G. Akemann, I.L. Aleiner, Y. Alhassid, A.
Altland, B.L.  Altshuler, M.E.  Berbenni-Bitsch, B.A. Berg, D.
Dalmazi, P.H.  Damgaard, Y.V.  Fyodorov, A.M. Garcia-Garcia, M.
G\"ockeler, T.  Guhr, M.\'A.  Hal\'asz, H. Hehl, A.D.  Jackson, N.
Kaiser, B. Klein, J.-Z. Ma, H.  Markum, S.  Meyer, S.M.  Nishigaki,
J.C.  Osborn, R.  Pullirsch, P.E.L. Rakow, A.  Sch\"afer, M.
Schnabel, A. Schwenk, B.  Seif, T.H.  Seligman, M.K.  \c Sener, B.D.
Simons, R.E.  Shrock, E.V.  Shuryak, A.V.  Smilga, M.A.  Stephanov, D.
Toublan, H.A.  Weidenm\"uller, T.  Wilke, and M.  Zirnbauer.  We also
thank I.L.  Aleiner for a critical reading of Sec.~\ref{otherfields}.

\end{document}